\documentclass{article}
\usepackage{a4wide}
\usepackage{algorithm}
\usepackage{algpseudocode}
\usepackage{amsmath}
\usepackage{amssymb}
\usepackage{authblk}
\usepackage{bbm}
\usepackage{graphicx}
\usepackage{hyperref}
\usepackage{setspace}
\usepackage{tabularx}
\usepackage{titling}

\graphicspath{{Figures/}}

\newcommand{\var}{\text{var}}
\newcommand{\cov}{\text{cov}}
\newcommand{\ve}[1]{{\bf #1}}
\newcommand{\dd}{\,\text{d}}
\DeclareMathOperator*{\argmax}{arg\,max}

\title{Measuring the accuracy of likelihood-free inference}
\author{Aden Forrow\thanks{\href{mailto:aden.forrow@maths.ox.ac.uk}{aden.forrow@maths.ox.ac.uk}} }
\author{Ruth E. Baker}
\affil{Mathematical Institute, University of Oxford}

\begin{document}
\maketitle
\abstract{
Complex scientific models where the likelihood cannot be evaluated present a challenge for statistical inference.
Over the past two decades, a wide range of algorithms have been proposed for learning parameters in computationally feasible ways, often under the heading of approximate Bayesian computation or likelihood-free inference.
There is, however, no consensus on how to rigorously evaluate the performance of these algorithms.
Here, we argue for scoring algorithms by the mean squared error in estimating expectations of functions with respect to the posterior.
We show that score implies common alternatives, including the acceptance rate and effective sample size, as limiting special cases.
We then derive asymptotically optimal distributions for choosing or sampling discrete or continuous simulation parameters, respectively.
Our recommendations differ significantly from guidelines based on alternative scores outside of their region of validity.
As an application, we show sequential Monte Carlo in this context can be made more accurate with no new samples by accepting particles from all rounds.
}

\section{Introduction}

A key challenge in modern Bayesian inference is developing algorithms to handle increasingly complex scientific data and models.
Standard approaches rely on explicit likelihood functions for either analytic calculation or, more commonly, Markov chain Monte Carlo (MCMC).
However, for many important models in contexts from cosmology~\cite{Alsing2019,Makinen2021} to psychology~\cite{Turner2014} and neuroscience~\cite{Fengler2021} computing the likelihood itself is intractable.
In such settings, inference must be done using simulations that sample from the unavailable likelihood.

Ongoing research over the past two decades has led to a wide range of algorithms for such \textit{likelihood-free inference}~\cite{Beaumont2009, Bonassi2015, DelMoral2012, Fearnhead2012, Fengler2021, Jarvenpaa2019, Lueckmann2018, Meeds2015, Papamakarios2016, Prescott2020, Price2018, Toni2009}, each aiming to perform accurate and efficient inference when the likelihood is only implicitly defined.
Thus far, however, there is no rigorously justified consensus on how to measure either accuracy or efficiency.
Proposed algorithms are accompanied by diverse methods for evaluating performance, which we refer to as \textit{scores} to avoid confusion with the technical meanings of \textit{metric} and \textit{measure}.
Examples include the acceptance rate~\cite{Bonassi2015,DelMoral2012, Filippi2013, Lintusaari2017,Price2018,Sisson2007, Sisson2018, Toni2009,Toni2009a}, effective sample size~\cite{Fearnhead2012, Prescott2021, Price2018,Sisson2020}, and precision in recovering known simulation parameters~\cite{Fengler2021,Lueckmann2017, Papamakarios2016,Price2018}.
This diversity of scores makes it difficult to compare algorithms, as the relationship between good performance on different scores is often unclear.

This paper argues for evaluating the accuracy of likelihood-free inference methods by the mean squared error (MSE) in approximating posterior expectations.
We begin by laying out the general likelihood-free inference problem in Section~\ref{s:problem_definition}.
In Section~\ref{s:score_descriptions}, we present the MSE score and the independent motivation for using it.
We then review commonly used alternative scores together with algorithms designed to optimize them, showing that each score either can be derived as an approximate special case of a function expectation or fails to account for important features of the match between true and approximate posteriors.

A key use for a reliable score is to compare the computational efficiency of different algorithms.
For many practical applications of likelihood-free inference, the dominant computational cost comes from performing expensive simulations from the model. 
We will ignore the complexity of other stages of an algorithm, though for sufficiently simple models they may be relevant.
The goal, then, is to optimize an appropriate score while minimizing the required number of simulations.
For this paper, we focus on optimizing over one aspect of likelihood-free inference, the choice of parameters with which to simulate the model.

We begin in section~\ref{s:discrete_parameters} with the case of discrete parameters, where we derive the set of simulations to run that optimizes the asymptotic MSE for a given function.
Our recommendations differ significantly from strategies proposed in the literature based on other scores.
Next, in Section~\ref{s:continuous_parameters}, we demonstrate that although continuous parameters present new challenges that make rigorous analysis difficult, the discrete results qualitatively translate to that setting. Here again, optimizing an algorithm for an inappropriate score can cause inefficiency.

\subsection{Likelihood-free inference}
\label{s:problem_definition}
In likelihood-free inference, we aim to use observed data $x^* \in \mathcal{X}$ to learn about the parameters $\theta \in \Theta$ governing a scientific model.
We approach the problem from a Bayesian perspective: we have a known prior distribution $p(\theta)$ encoding our initial knowledge about $\theta$ and seek to infer a posterior distribution $p(\theta|x^*)$, which is related to the likelihood $p(x^*|\theta)$ and prior via Bayes' Theorem:
\begin{equation}
p(\theta | x^*) = \frac{p(x^* | \theta) p(\theta)}{p(x^*)}.
\label{e:bayes}
\end{equation}

Unlike in traditional Bayesian inference, we do not assume $p(x^*|\theta)$ can be computed, even up to a normalizing constant.
Such intractable likelihoods often occur when a model includes many unobserved latent parameters. In a stochastic epidemic model, for example, explicitly calculating the likelihood of a certain number of patients testing positive for the disease may require summing over all possible configurations of asymptomatic or unconfirmed cases.
To avoid that impossibly expensive computation, we learn about the likelihood by running simulations to sample from $p(x|\theta)$.

Our goal is to create an approximation $\hat p(\theta|x^*)$ to the posterior.
Any algorithm to do so must implement two steps:
\begin{enumerate}
\item Choose $\{\theta_i\}$ and generate simulated data $\{x_i\}$.
\item Estimate $\hat p(\theta|x^*)$ from the samples $\{(\theta_i, x_i)\}$.
\end{enumerate}
The steps may not be separable from each other: many algorithms~\cite{Beaumont2009,Filippi2013,Papamakarios2019,Sisson2007} use an intermediate posterior approximation to guide subsequent choices of simulation parameters.
In addition, the form of $\hat p(\theta|x^*)$ varies depending on the algorithm: it could be, for example, a set of samples approximately from $p(\theta|x^*)$~\cite{Filippi2013} or a mixture density network~\cite{Papamakarios2016} mapping $\theta$ to $p(\theta|x^*)$.
We assume throughout that $x_i$ is drawn from $p(x|\theta_i)$, though alternatives with approximate models are possible~\cite{Prescott2020, Prescott2021}.

A popular strain of likelihood-free inference is \textit{Approximate Bayesian computation} (ABC), whose most basic form approximates the posterior with the empirical distribution of $\{(\theta_i, x_i)\}$
weighted by a kernel $K\left(\Delta(x_i, x^*)/\epsilon\right)$ dependent on some distance function $\Delta$ and threshold $\epsilon$.
ABC involves several important complications we will not analyze here, including the choice of $\Delta$, often involving summary statistics, and the setting of $\epsilon$.
For our theoretical results in Sections~\ref{s:discrete_parameters} and~\ref{s:continuous_parameters}, we assume that the probability of observing $x^*$ is nonzero for some $\theta$.
If $\mathcal{X}$ is discrete, that assumption is trivial; if the raw data $y^*$ is continuous, we could let $x^*$ be the observation that $y$ is within an $\epsilon$-ball of $y^*$, similarly to standard ABC.

\begin{table}
\begin{center}
\begin{tabularx}{1\linewidth}{l X}
Symbol & Meaning\\
\hline
$x^*$ & Observed data\\
$x$ & Generic element of data space $\mathcal{X}$\\
$N$ & Total number of simulations performed \\
$N_{acc}$ & Number of samples $(\theta_i, x_i)$ accepted, if applicable\\
$\bar f$ & $\mathbb{E}_{p(\theta|x^*)}\left[f(\theta)\right]$, expectation of $f(\theta)$ with respect to true posterior\\
$q(\theta)$ & Unspecified probability distribution over $\theta$, often an importance distribution\\
$\hat q(\theta)$ & An estimate of $q(\theta)$
\end{tabularx}
\end{center}
\caption{Notation used throughout the paper.}
\end{table}

\subsection{Scoring algorithm performance}
\label{s:score_descriptions}

\begin{table}
\begin{center}
\begin{tabularx}{1\linewidth}{X X X}
Evaluation method & Motivation & LFI references\\
\hline
Expectation MSE & Part of IPM, explicitly depends on function of interest & \cite{Barber2015, Li2018, Prescott2020} \\
Integral probability metrics \cite{Sriperumbudur2009} & Standard class of metrics on probability distributions &(TV) \cite{Lueckmann2018} (MMD) \cite{Papamakarios2019} $\qquad$ (CDF) \cite{Warne2018}\\
Acceptance rate & Easy to check, sample size for rejection sampling &  \cite{Bonassi2015,DelMoral2012, Filippi2013, Lintusaari2017,Price2018,Sisson2007, Sisson2018, Toni2009,Toni2009a}\\
Effective sample size \cite{Elvira2018, Liu1996} & Adjusting acceptance rate for sample weights & \cite{Fearnhead2012, Meeds2015, Prescott2021, Price2018,Sisson2020} \\
Normalized posterior MSE & Natural metric on distributions &  \cite{Biau2015, Blum2010}\\
Unnormalized posterior MSE & Normalization difficult to analyze & \cite{Jarvenpaa2019} \\
$\phi$-divergences \cite{Sriperumbudur2009} & Standard divergences between probability distributions & (KL) \cite{Beaumont2009, Filippi2013, Lintusaari2017, Papamakarios2016}, (TV) \cite{Lueckmann2018}\\
Log-likelihood accuracy & Penalizes misestimating likelihood as zero &  \cite{VanOpheusden2020}\\
Null hypothesis significance tests & Traditional statistical approach & \cite{Turner2014}\\
Distribution plots & Visualization shows hard-to-summarize features & \cite{Lueckmann2017, Meeds2015, Papamakarios2016,Prangle2019,Price2018,Sisson2007,Turner2014}\\
Posterior probability of true parameters & True posterior not needed & \cite{Fengler2021,Lueckmann2017, Papamakarios2016, Papamakarios2019, Price2018} \\
Posterior predictive checks & Easy to check, diagnoses significant failures &\cite{Lueckmann2017} \\
\end{tabularx}
\end{center}
\caption{Summary of scores proposed in the literature. The third column is a not-exhaustive list of references that use or recommend optimizing for each score.}
\end{table}

\subsubsection{The mean squared error score $\mathcal{S}_{d_f}$}

A first step in evaluating an algorithm's performance is to choose a quantitative discrepancy function $d(\hat p(\theta|x^*), p(\theta|x^*))$ describing how well $\hat p (\theta |x^*)$ approximates the posterior.
For any such function, we can take an expectation over the randomness in computing $\hat p (\theta |x^*)$ to form a score for the algorithm,
\begin{equation}
\mathcal{S}_d = \mathbb{E}\left[d(\hat p(\theta|x^*) ,  p(\theta|x^*) )\right],
\label{e:generic_score}
\end{equation}
which we aim to minimize.
The function $d$ should target features we care about in the match between $\hat p(\theta|x^*)$ and $p(\theta|x^*)$.

Many relevant features are encoded in expectations of functions. The posterior mean is the expectation of $\theta$; in a model selection problem, the posterior probability of model $M$ is the expectation of an indicator function for $M$; $\mathcal{I}$ is a posterior credible interval covering probability $P$ if $P$ is the posterior expectation of the indicator $\mathbbm{1}(\theta \in \mathcal{I})$.
Estimation of both the mean and second moment enables estimation of the posterior variance, which is a common goal~\cite{Fengler2021, Lueckmann2017}.

We can assess the accuracy of estimation of such features by choosing a function $f(\theta)$, or set of functions collected into a vector, and setting
\begin{equation}
d_f (\hat q ,  q ) = \left\| \mathbb{E}_{\hat q}\left[ f(\theta)\right] -  \mathbb{E}_{ q}\left[ f(\theta)\right]\right\|^2,
\label{e:MSE_discrepancy}
\end{equation}
where we choose the squared error for later analytical convenience.
The expectations in Eq.~\ref{e:MSE_discrepancy} are taken with respect to $\hat q(\theta)$ and $q(\theta)$, i.e. the approximate and true posteriors, in contrast to Eq.~\ref{e:generic_score} where the expectation is over the distribution of approximate posteriors produced by an algorithm.
The corresponding score $\mathcal{S}_{d_f}$ is common in the literature, used by
Barber et al.~\cite{Barber2015} to evaluate rates of convergence of rejection sampling, 
Li and Fearnhead~\cite{Li2018} to argue sampling simulation parameters from either prior or posterior is asymptotically inefficient, and Prescott and Baker~\cite{Prescott2020} to evaluate the efficiency of multifidelity ABC.

While $d_f(\hat q, q) = 0$ for an individual function $f$ does not alone imply $\hat q = q$, 
the discrepancy function $d_f$ can be extended to a metric on probability distributions by taking a square root and a supremum over functions in a sufficiently rich class $\mathcal{F}$ of bounded, measurable, real-valued functions~\cite{Sriperumbudur2009}:
\begin{equation}
d_{\mathcal{F}} (\hat q ,  q ) = \sup_{f\in \mathcal{F}} \bigg| \mathbb{E}_{\hat q}\left[ f(\theta)\right] -  \mathbb{E}_{ q}\left[ f(\theta)\right]\bigg|.
\label{e:IPM}
\end{equation}
Such metrics, called integral probability metrics (IPMs), include many standard metrics on probability distributions. 
For example, choosing $\mathcal{F} = \{ f : \sup_\theta |f(\theta)| \le 1\}$ yields the total variation (TV) distance~\cite{Sriperumbudur2009}; choosing the class of functions with Lipschitz constant at most 1 yields the Wasserstein distance~\cite{Sriperumbudur2009}; and choosing the unit ball in a reproducing kernel Hilbert space yields the maximum mean discrepancy (MMD)~\cite{Gretton2012}.

The core of an IPM is the error in expectations of individual functions. By definition, $d_{\mathcal{F}}(\hat q, q)$ will be small if and only if $d_f(\hat q, q)$ is small for every $f$ in $\mathcal{F}$.
We see this connection to a proper metric as a motivation for using the MSE score $\mathcal{S}_{d_f}$, which we focus on over IPMs for three reasons.
First, dealing with the supremum in Eq.~\ref{e:IPM} may be analytically difficult.
Second, if $\mathcal{F}$ is too rich then a small IPM distance $d_\mathcal{F}\left(\hat p(\theta|x^*), p(\theta|x^*)\right)$ is too stringent a goal.
For example, if we approximate a continuous posterior in $D \ge 2$ dimensions with the empirical distribution of $N$ samples, the total variation distance between $\hat p(\theta|x^*)$ and $p(\theta|x^*)$ is always 1 and the Wasserstein distance decays as $N^{-1/D}$~\cite{Weed2017}, both much worse than the $N^{-1/2}$ convergence rate for any particular $f(\theta)$.
Finally, considering each $f$ individually forces us to pay attention to which expectations we estimate well or poorly, which enables improved efficiency for specific functions of interest.

The previous paragraphs demonstrate that $\mathcal{S}_{d_f}$ is a plausible candidate score.
The next stage of our argument for using it is to show that several commonly used alternatives are, at least approximately, equivalent to choosing a specific function $f(\theta)$ and, in some cases, a specific inference algorithm.
We first discuss the \textit{acceptance rate} and \textit{effective sample size}, in the process introducing implementations of likelihood-free inference where those scores are often considered.
We then show how both $\phi$-divergences~\cite{Sriperumbudur2009} and the MSE in $p(\theta|x^*)$ imply particular targets $f(\theta)$.
Finally, we conclude our section on scores by mentioning the remaining approaches we have seen for evaluating posterior approximations, which may complement the MSE score without replacing it as a goal.

\subsubsection{Scores connected to $\mathcal{S}_{d_f}$}

A traditional score for ABC is the acceptance rate, which comes from the simplest ABC variant, \textit{rejection sampling} (Algorithm~\ref{alg:ABC_rejection}). Simulation parameters $\{\theta_i\}$ are sampled independently from the prior and \textit{accepted} if the simulation output $x_i$ is sufficiently close to the observed data.
The posterior is approximated as the empirical distribution of accepted parameters, or, equivalently, by giving each accepted parameter weight $w_i = 1$ and each rejected parameter weight $w_i = 0$.
For the purposes of this paper, except when discussing sequential Monte Carlo, ``sufficiently close'' will be equality $x_i = x^*$,
in which case the rejection sampling output $\{\theta_i\}$ is a set of $N_{acc}$ independent samples from the posterior.
We assume $x_i = x^*$ occurs with nonzero probability to avoid complications from either setting a threshold $\epsilon$ on $\|x_i - x^*\|$ or making assumptions about the structure of $p(x|\theta)$.

\begin{algorithm}
\caption{ABC rejection sampling}\label{alg:ABC_rejection}
\begin{algorithmic}
\For{$i = 1:N$}
    \State $\theta_i \sim p(\theta)$
    \State $x_i \sim p(x|\theta_i)$
	\If{$x_i = x^*$}
			\State $w_i \gets 1$
	\Else
			\State $w_i \gets 0$
	\EndIf
\EndFor
\\
\Return $\{\theta_i : w_i = 1\}$
\end{algorithmic}
\end{algorithm}

We can compute an approximate posterior expectation of any function $f(\theta)$ as the empirical average $\hat {\bar f} = N_{acc}^{-1}\sum_{i=1}^{N_{acc}} f(\theta_i)$ over the accepted samples.
Because the samples from Algorithm~\ref{alg:ABC_rejection} are independent, $\hat {\bar f}$ is unbiased and has variance $N_{acc}^{-1}$ times the posterior variance of $f(\theta)$.
For rejection sampling, then, the \textit{acceptance rate} $N_{acc}/N$ is a proxy for the MSE score $\mathcal{S}_{d_f}$ with any function.

Clearly, too small an acceptance rate makes inference difficult.
If the expected acceptance rate is not sufficiently large compared to $1/N$, there is a risk that no samples will be accepted and no information gained about the posterior.
For realistic problems, the acceptance rate of Algorithm~\ref{alg:ABC_rejection} may be vanishingly small if the prior covers large regions of parameter space where the likelihood of generating data similar to $x^*$ is tiny.

This issue leads to a standard argument against rejection sampling: it is inefficient because too many samples are rejected~\cite{Bonassi2015,Lintusaari2017,Papamakarios2016,Sisson2018,Sisson2007,Toni2009}. Implicitly, that argument suggests that algorithms with higher acceptance rates should systematically outperform algorithms with lower acceptance rates.
In that spirit, the acceptance rate is included as an algorithm quality score in a wide range of papers~\cite{Bonassi2015,DelMoral2012, Filippi2013, Price2018, Sisson2018, Sisson2007, Toni2009a}, in some cases together with other evaluations.

A long strain of research~\cite{DelMoral2012,Filippi2013,Sisson2007,Toni2009a} combines Algorithm~\ref{alg:ABC_rejection} with importance sampling to improve the acceptance rate.
Rather than sampling from the prior, $\{\theta_i\}$ are drawn from an \textit{importance distribution}  $q(\theta)$ and, if accepted, given a weight $w_i = p(\theta_i)/q(\theta_i)$.
The resulting weighted empirical distribution $\hat p(\theta|x^*)$, with weights normalized to sum to one, approximates the posterior. Expectations can again be approximated with empirical averages:
\begin{equation}
\label{e:rejection_sampling_expectation}
\mathbb{E}_{\hat p(\theta|x^*)} [f(\theta) ] = \frac{\sum_{i=1}^N w_i f(\theta_i)}{\sum_{i=1}^N w_i}.
\end{equation}
This estimate is biased due to the ratio, though ideally not significantly biased. In some importance sampling contexts~\cite{Elvira2018} the expectation of the denominator can be explicitly calculated to give an unbiased estimator; here, that is not possible because it depends on the unknown likelihood:
\begin{align}
\mathbb{E}\left[\sum_{i=1}^N w_i\right] 
 = N \sum_\theta p(x^* | \theta) \frac{p(\theta)}{q(\theta)} q(\theta)
 = N p(x^*) = \mathbb{E}\left[N_{acc}\right].
\end{align}

The importance sampling variant leads naturally to a combination of ABC with \textit{sequential Monte Carlo} (SMC)~\cite{DelMoral2012, Filippi2013,Sisson2007, Toni2009}. ABC-SMC methods do multiple rounds of importance sampling from an adaptively-constructed approximation to the posterior. In the first round, the $\theta_i$ are sampled from the prior and accepted if $x_i$ is within a relatively large tolerance $\epsilon_1$ of $x^*$. In each subsequent round $k$, parameters, called \textit{particles}, are perturbed samples from the previous round's accepted particles and the tolerance $\epsilon_k$ is reduced. 
The traditional algorithm output is the weighted set of particles from the final round, though including particles from all rounds may significantly improve accuracy (Supplement Section~\ref{app:ABC-SMC_keeping_all_rounds}).
The goal of running multiple rounds is to get as close as possible to sampling parameters from the posterior, which has been claimed to give ``the maximum possible efficiency of ABC samplers'' \cite{Sisson2007}.

However, despite its intuitive appeal, sampling $\{\theta_i\}$ from the posterior may not improve efficiency.
The change from sampling parameters from $p(\theta)$ to sampling parameters from the importance distribution $q(\theta)$ broke the direct connection between acceptance rate and estimator variance: the variance of $\hat {\bar f}$ is no longer $N_{acc}^{-1}$ times the posterior variance of $f(\theta)$.
In general, outside of the regime where tiny acceptance rates make inference infeasible, the relationship between the acceptance rate and other accuracy scores is not well understood. 
Truly maximizing the acceptance rate would be undesirable if it were possible, as all simulations would be at the maximum likelihood parameter values and give no information about the rest of the posterior.

A key factor ignored in focusing solely on the acceptance rate is variability in the sample weights.
The effective sample size is a heuristic designed to account for that variability.
It is defined for a set of $N$ samples $\{\theta_i\}$ with weights $\{w_i\}$ as
\begin{equation}
\widehat{ESS} = \frac{\left(\sum_{i=1}^N w_i\right)^2}{\sum_{i=1}^N w_i^2}.
\label{e:ESS}
\end{equation}
The effective sample size $\widehat{ESS}$ can be derived through a series of approximations to the variance of an importance sampling estimate of $\mathbb{E}_{p(\theta|x^*)}\left[f(\theta)\right]$~\cite{Elvira2018, Liu1996} relative to the variance of $f(\theta)$ with $\theta\sim p(\theta|x^*)$. 
In contexts outside this paper, the term effective sample size sometimes refers directly to the ratio of the variance of an estimator to the variance of a single posterior sample, which may be estimated differently~\cite{Vehtari2021}.
Here we will only consider the estimate in Eq.~\ref{e:ESS}, where the hat emphasizes that we do not have the true relative variance. 
$\widehat{ESS}$ ranges from 1 to $N$ and is larger when the weights are more uniform. 
For rejection sampling, where $w_i = 1$ for accepted samples and zero for rejected samples, the effective sample size is always the number of accepted samples, $N_{acc}$. 

Despite criticism~\cite{Elvira2018} in the context of importance sampling, where it was derived~\cite{Liu1996}, $\widehat{ESS}$ is commonly used to measure sample quality.
Fearnhead and Prangle~\cite{Fearnhead2012} proposed an importance distribution designed to maximize $\widehat{ESS}$; Del Moral et al.~\cite{DelMoral2012} and Prangle~\cite{Prangle2019} use observed $\widehat{ESS}$ to control the rate at which target distributions approach the posterior; and many authors~\cite{Meeds2015, Prescott2021, Price2018, Sisson2020} use it to evaluate their methods.
The clearest advantage of Eq.~\ref{e:ESS} is that it is easy to compute without information about the function of interest $f(\theta)$ or the true posterior.

One difficulty with either the acceptance rate or the effective sample size as general-purpose scores is that they are not defined for all algorithms.
We could not use $\widehat{ESS}$, for example, to compare an algorithm using Eq.~\ref{e:rejection_sampling_expectation} to one that integrated a kernel regression estimate (Supplement Section~\ref{app:kernel_regression}) based on the same set of weighted samples $\{\theta_i, x_i,w_i\}$.

More importantly, the approximations used to derive Eq.~\ref{e:ESS} may not apply for any particular function.
The derivation in~\cite{Liu1996}, done for pure importance sampling rather than likelihood-free inference, leads to $\widehat{ESS}$ by neglecting an error term $\mathbb{E}\left[ \left(w - \mathbb{E}[w]\right) \left\|f(\theta) - \mathbb{E}[f(\theta)]\right\|^2\right]$, where all expectations are taken with respect to the importance sampling target.
The error is straightforwardly zero in two cases: rejection sampling, because $w - \mathbb{E}[w]$ is zero everywhere; and target functions where $ \left\|f(\theta) - \mathbb{E}[f(\theta)]\right\|^2$ is constant, which if $f(\theta)\in \mathbb{R}$ are affine transformations of indicator functions for $50\%$ credible intervals.
We are not aware of any arguments for why it should be small in general.

The independence of $\widehat{ESS}$ from the target $f(\theta)$ is therefore an illusion. While Eq.~\ref{e:ESS} does not incorporate $f(\theta)$, once we have moved away from Algorithm~\ref{alg:ABC_rejection} the degree to which it is an accurate approximation of the estimator variance does depend on $f(\theta)$.
Moreover, the derivation of $\widehat{ESS}$ assumed the $\theta_i$ were sampled independently; choosing $\theta_i$ without sampling~(\cite{Jarvenpaa2019} and Section~\ref{s:discrete_parameters}), or with stratified sampling (Supplement Section~\ref{app:stratified}) changes the variance.
Optimizing $\widehat{ESS}$ despite its inaccuracy then leads to suboptimal estimator variance, as we will see in Sections~\ref{s:discrete_parameters} and~\ref{s:continuous_parameters} after we complete our survey of possible scores.

If we are strongly opposed to choosing target functions, a tempting alternative is to consider the value of the posterior itself.
We can either pick one $\theta^*$ and set
\begin{equation}
d_{post, \theta^* } (\hat q ,  q )= \left(\hat q (\theta^*) - q (\theta^*)\right)^2
\end{equation}
or integrate to
\begin{equation}
d_{post}(\hat q ,  q ) = \int d_{post, \theta^*} (\hat q, q ) \dd \theta^*.
\end{equation}

Biau et al.~\cite{Biau2015} computed the rate of convergence of $\mathcal{S}_{d_{post}}$ for ABC algorithms where the acceptance rate rather than the acceptance threshold $\epsilon$ on $\|x_i - x^*\|$ is fixed, 
while Blum~\cite{Blum2010} calculated the asymptotic contributions of bias and variance to $\mathcal{S}_{d_{post, \theta^*}}$ for a kernel regression estimate $\hat p(\theta^*|x^*)$.
Both assumed simulation parameters were sampled from the prior.

Rather than being independent of $f$, however, both options are special cases of the MSE score. $S_{d_{post, \theta^* }}$ and $S_{d_{post}}$ are equal to $S_{d_f}$ with $f_{post, \theta^*}(\theta) = \delta(\theta^* - \theta)$ and $f_{post}(\theta) = \delta(\theta' - \theta)$ respectively, where the latter is considered as a function of $\theta'$.

A minor variant of $S_{d_{post}}$ is to target the unnormalized posterior $p(x^*|\theta)p(\theta)$, as done by J\"arvenp\"a\"a et al.~\cite{Jarvenpaa2019}.
This simplification is computationally convenient, as dealing with normalization complicates calculations.
It is, however, a simplification, and ignores the effect of error in estimating the normalizing constant on estimating $p(\theta|x^*)$.
We will see an example in Section~\ref{s:discrete_parameters} where that error dominates.

Our final class of scores to connect to the MSE score are $\phi$-divergences~\cite{Sriperumbudur2009}. Each is defined by a convex function $\phi$ with $\phi(1) = 0$ as
\begin{equation}
d_{\phi\text{-div}}(\hat q,q) = \int \phi\left(\frac{\hat q(\theta)}{q(\theta)}\right) q(\theta) \dd\theta
\label{e:phi_divergence}
\end{equation}
This includes the Kullback-Leibler (KL) divergence $KL(\hat q|q)$, with $\phi(t) = t \log t$, as well as the TV distance, with $\phi(t) = \frac{1}{2}|t -1|$.
Certain ABC-SMC algorithms~\cite{Beaumont2009, Filippi2013} aim to minimize the KL divergence between the joint distributions of particles of consecutive iterations and independent copies of the posterior. Other authors~\cite{Lintusaari2017, Papamakarios2016} use KL divergences for final evaluations.

With its connections to information theory, the KL divergence is a compelling candidate for comparing the true and approximate posteriors.
However, in the relevant limiting case of an accurate algorithm $\phi$-divergences may be approximated by a difference of function expectations, as we now show. 
If $\phi$ is twice continuously differentiable at 1, the integrand of Eq.~\ref{e:phi_divergence} can be expanded as
\begin{equation}
\phi\left(\frac{\hat q(\theta)}{q(\theta)}\right) q(\theta)
= \phi(1)q(\theta) + \phi'(1) q(\theta) \left(\frac{\hat q(\theta)}{q(\theta)} - 1\right)
+ \frac{1}{2}\phi''(1) q(\theta)\left(\frac{\hat q(\theta)}{q(\theta)} - 1\right)^2
+ O\left(\left(\frac{\hat q(\theta)}{q(\theta)} -1 \right)^3 \right)
\end{equation}
The first term is zero by definition; the second term integrates to zero when $\hat q $ and $q$ are both normalized. Suppose  $\hat q$ is a close enough approximation to $q$ that the last term is negligible, which for likelihood-free inference means that we have used enough computational effort to get a good approximation to the posterior. Then
\begin{equation}
d_{\phi\text{-div}}(\hat q,q) \approx \int \frac{\phi''(1)}{2} \frac{\left(\hat q(\theta) - q(\theta)\right)^2}{q(\theta)} \dd\theta,
\label{e:phi_divergence_approximation}
\end{equation}
which is equal to the expectation squared error $d_f(\hat q, q)$ with $f(\theta) = \left(\phi''(1)/ \left(2q(\theta)\right)\right)^{1/2}\delta(\theta' - \theta)$. A similar argument leads to the same result for the leading order term in $d_{\phi\text{-div}}(q, \hat q)$ (Supplement Section~\ref{app:phi_divergence}).

 In the context of a consistent likelihood-free inference algorithm, therefore, the KL divergence  between $\hat p(\theta|x^*)$ and $p(\theta|x^*)$, as well as similar $\phi$-divergences, will behave similarly to the MSE of a data-dependent function emphasizing parameters with low posterior probability.
We validate the approximation for a simple example in Fig.~\ref{f:two_sample_MSE_simulations}.
The only commonly used $\phi$ divergence which is not smooth, and so not covered by this argument, is the TV distance which can be defined independently in terms of function expectations.

For these last scores $\mathcal{S}_{d_{post}}$ and $\mathcal{S}_{d_{\phi\text{-div}}}$, as for TV and Wasserstein distances, we run into complications with discrete approximations to continuous posteriors.
Both $KL\left(\hat p(\theta|x^*)|p(\theta|x^*)\right)$ and $d_{post}\left(\hat p(\theta|x^*),p(\theta|x^*)\right)$ are infinite if $\hat p(\theta|x^*)$ is an empirical distribution of samples.
While this problem is not insurmountable, as an algorithm that returned posterior samples could be supplemented with a kernel regression estimate or equivalent procedure, choosing the MSE score $\mathcal{S}_{d_f}$ avoids it.

For quantitative evaluation of the accuracy of $\hat p(\theta|x^*)$, then, we recommend choosing functions of interest and estimating $\mathcal{S}_{d_f}$.
Convergence in stricter metrics like TV, though valuable if achievable, likely isn't necessary for many applications.
Meanwhile, other commonly used quantitative scores imply a specific target function, a specific algorithm structure, or both.
None of them are truly independent of the target.
Using a different score without explicitly accounting for $f(\theta)$ merely hides the dependence.

\subsubsection{Alternative evaluation methods}

Alternative ways to compare $\hat p(\theta|x^*)$ and $p(\theta|x^*)$ not directly connected to the MSE score do exist.
For completeness, we first mention two less commonly used options.
Van Opheusden et al.~\cite{VanOpheusden2020} propose inverse binomial sampling (IBS) as a way to get an unbiased estimate of the log-likelihood of the model, rather than the likelihood itself. 
Estimating log-likelihoods is more difficult than estimating likelihoods: small errors in small likelihoods are magnified because $\log\left(p(x^*|\theta)\right)$ diverges as the likelihood goes to zero.
The method therefore spends most  computational effort on low-likelihood regions of parameter space to distinguish between small and very small likelihoods.
Often, however, the difference between small and very small does not matter for conclusions that will be drawn from $\hat p(\theta|x^*)$.
In cases where there is a particular low-probability event of interest, it can be targeted with a specific $f(\theta)$.

Separately, Turner and Sederberg~\cite{Turner2014} used a Kolmogorov-Smirnov (K-S) test for whether one of their true and approximate posterior pairs were the same.
Such an approach combines the controversial aspects of null hypothesis significance testing~\cite{Gelman2013,Wasserstein2016} with technical difficulties like choosing the sample size in the K-S test.

The alternative evaluation methods we find more promising are qualitative rather than quantitative.
These aim only to check whether $\hat p (\theta|x^*)$ is very wrong.
Perhaps the simplest approach is to plot $\hat p(\theta|x^*)$ together with $p(\theta|x^*)$ for an example where the true posterior is known.
This is ubiquitous in the literature~\cite{Lueckmann2017, Meeds2015, Papamakarios2016,Prangle2019,Price2018,Sisson2007,Turner2014}, for good reason.
Plotting is easy to implement and makes the most important differences between $\hat p(\theta|x^*)$ and $p(\theta|x^*)$ visually obvious, without requiring the user to decide beforehand what features they care about.

Rather than plotting the full distribution, it is also common~\cite{Fengler2021,Lueckmann2017, Papamakarios2016}  to compare the posterior with the ground truth parameters of synthetic data.
Such a comparison does not require the true posterior to be tractable.
A good estimate by this criterion concentrates on the true value, or assigns high probability to the true value~\cite{Papamakarios2019}, or has mean near the true value~\cite{Lintusaari2017}.
Like plotting the full posterior, comparing to a known ground truth is easy to do and can diagnose significant algorithm failures.
It does not require knowing the posterior and can be supplemented with calibration tests~\cite{Papamakarios2019} or cross-validation~\cite{Price2018}.
Alone, however, ground truth comparison provides limited information. A distribution may have high density at the ground truth parameters while still being a poor approximation to the true posterior, either by underestimating uncertainty or because the data is atypical for the true parameters.

Neither of the previous approaches is possible when both the true posterior and the true parameters are unknown.
In that case, it is still possible to sample parameters from the approximate posterior, simulate from them, and compare the simulation results to the data.
Like posterior plots, such checks can diagnose significant algorithm failures where the posterior predictive distribution is far from the data.
Good performance on a posterior predictive check, however, does not imply that the approximate posterior has the right level of uncertainty.
An approximate posterior that puts too little weight on plausible alternative parameter values would still perform well.

In general, qualitative methods are useful but insufficient for comparing algorithms that all give plausible results.
We would like to be able to evaluate the accuracy and computational efficiency of the many proposed likelihood-free inference algorithms whose approximate posteriors for standard examples are not obviously unreasonable.
Of the quantitative scores we considered, $\mathcal{S}_{d_f}$ is the most generic and broadly applicable.
With that choice, we are now ready to compare algorithms' efficiency.


\section{Optimally choosing discrete parameters  }
\label{s:discrete_parameters}

Once we have a satisfactory score, a natural followup is to ask what algorithm design optimizes accuracy for a fixed computational cost.
To highlight the effect of the choice of score on the answer to that question, we first consider the simplified case where the set of parameters is discrete. This could be, for example, a model selection problem where each model has no internal parameters to be estimated.
As we outlined in Section~\ref{s:problem_definition}, likelihood-free inference algorithms have two parts: choosing parameters for model simulations, and estimating the posterior based on the simulation results.
With discrete parameters, the second part will be trivial, allowing us to focus on how parameters should be chosen.

Let $\Theta = \{\theta_i\}$ for $i = 1, 2, \ldots, k$. 
We will let $N$, the number of simulations performed, go to infinity while $k$, the number of possible parameter values, remains fixed and finite.
For each $i$, we choose a number $n_i$ of simulations to run with $\theta=\theta_i$; of these, $n_i^*$ yield $x = x^*$.
We assume $n_i$ and $n_i^*$ are sufficiently large that no estimator significantly improves 
 on the maximum likelihood estimate $\hat p_i^* = n_i^*/n_i$ of the likelihood $p_i^* = p(x^*|\theta_i)$. Let $\pi_i = p(\theta_i)$ be the prior on parameters.
Importantly, $n_i/N$ is the deterministic fraction of simulations with $\theta_i$, not the probability of sampling $\theta_i$ for simulation. Randomness in selecting $\theta_i$ would add variance, as discussed in the supplement (Section~\ref{app:sampling_parameters}).

Given a function $f(\theta)$, the resulting estimate of the true posterior expectation $\bar f$ is
\begin{align}
\mathbb{E}_{\hat p (\theta|x^*)}[f(\theta)] &= \frac{\sum_{i=1}^k f(\theta_i)\hat p_i^* \pi_i}{\sum_{i=1}^k \hat p_i^* \pi_i} \equiv \frac{R}{S}.
\end{align}
The numerator and denominator are separately unbiased estimates of $\mu_R = p(x^*) \bar f$ and $\mu_S = p(x^*)$ respectively; their ratio has bias of order $1/n_i$, which will be asymptotically negligible in comparison to the variance (Supplement Section~\ref{app:delta_method}).

Using the delta method (Supplement Section~\ref{app:delta_method}), the variance can be approximated as
\begin{align}
\var\left(\frac{R}{S}\right)
& \approx \frac{1}{\mu_S^2} \var(R) -2\frac{\mu_R}{\mu_S^3}\cov(R, S) + \frac{\mu_R^2}{\mu_S^4}\var(S)+ O\left(n_i^{-3/2}\right).
\label{e:delta_method_variance}
\end{align}
Each term in Eq.~\ref{e:delta_method_variance} can be calculated explicitly:
\begin{align}
\var(R) &= \sum_i f(\theta_i)^2\pi_i^2 \var(\hat p_i^*) = \sum_{i=1}^k f(\theta_i)^2\pi_i^2  p_i^*(1 - p_i^*)/n_i ;\\
\cov(R, S) &= \sum_i f(\theta_i)\pi_i^2 \var(\hat p_i^*) = \sum_{i=1}^k f(\theta_i)\pi_i^2  p_i^*(1 - p_i^*)/n_i;
\\
\var(S) &= \sum_i \pi_i^2 \var(\hat p_i^*) = \sum_{i=1}^k \pi_i^2  p_i^*(1 - p_i^*)/n_i.
\end{align}
This leads, after some algebra, to an expression for the asymptotic variance:
\begin{align}
\var\left(\frac{R}{S}\right)
& =  \mu_S^{-2}\sum_{i=1}^k \pi_i^2 \frac{p_i^*(1-p_i^*)}{n_i}\left(
 f(\theta_i) - \bar f
\right)^2 + O\left(n_i^{-3/2}\right).
\label{e:discrete_asymptotic_variance}
\end{align}

The only choice in our algorithm is how to select the $n_i$ with a constrained total simulation budget $N = \sum_{i=1}^k n_i$.
From a calculation with a Lagrange multiplier (Supplement Section~\ref{app:optimal_MSE_distribution}), the $n_i$ that minimize variance satisfy
\begin{equation}
n_i \propto \pi_i  \left(p_i^*(1-p_i^*)
\right)^{1/2} \left|
f(\theta_i) - \bar f
\right|.
\label{e:discrete_optimal_n}
\end{equation}
This involves three factors. First, $\pi_i$: we should simulate more from regions consistent with our prior knowledge. Second, $\left(p_i^*(1-p_i^*)
\right)^{1/2}$, the standard deviation of the unknown likelihood: we should spend more effort on regions where the likelihood is harder to learn.
Third, $\left| f(\theta_i) - \bar f \right|$, the difference between $f(\theta_i)$ and the true posterior mean, $\bar f$: we should concentrate on regions where a changed estimate of the likelihood would have a greater effect on the estimate of $\bar f$. 
As in the case of importance sampling~\cite{Elvira2018}, the dependence on $f$ is unavoidable. The optimal simulation strategy for one function may be far from optimal for another.

In the case $k=2$, Equation~\ref{e:discrete_optimal_n} can be instructively simplified: 
\begin{align}
n_1 &\propto \pi_1  \left(p_1^*(1-p_1^*)
\right)^{1/2} \left|
f(\theta_1) - \frac{f(\theta_1)p_1^* \pi_1 + f(\theta_2)p_2^*\pi_2}{ p_1^* \pi_1 + p_2^*\pi_2}
\right|\\
& = \pi_1  \left(p_1^*(1-p_1^*)
\right)^{1/2}\frac{p_2^* \pi_2}{p_1^* \pi_1 + p_2^*\pi_2} \left| f(\theta_1) - f(\theta_2)\right|
\\
& \propto \left(p_2^* (1 - p_1^*)\right)^{1/2},
\end{align}
where in the last line we removed all factors that are shared between the expressions for $n_1$ and $n_2$.
Both the prior and the function $f$ have disappeared. The optimal algorithm depends on neither. Moreover, both remaining factors imply that increasing $p_1^*$ decreases the optimal $n_1$. 
To minimize estimation error for this maximally simplified problem, you should always spend more computational effort on the parameter value the data support less.

\begin{figure}
\includegraphics[width=1\linewidth]{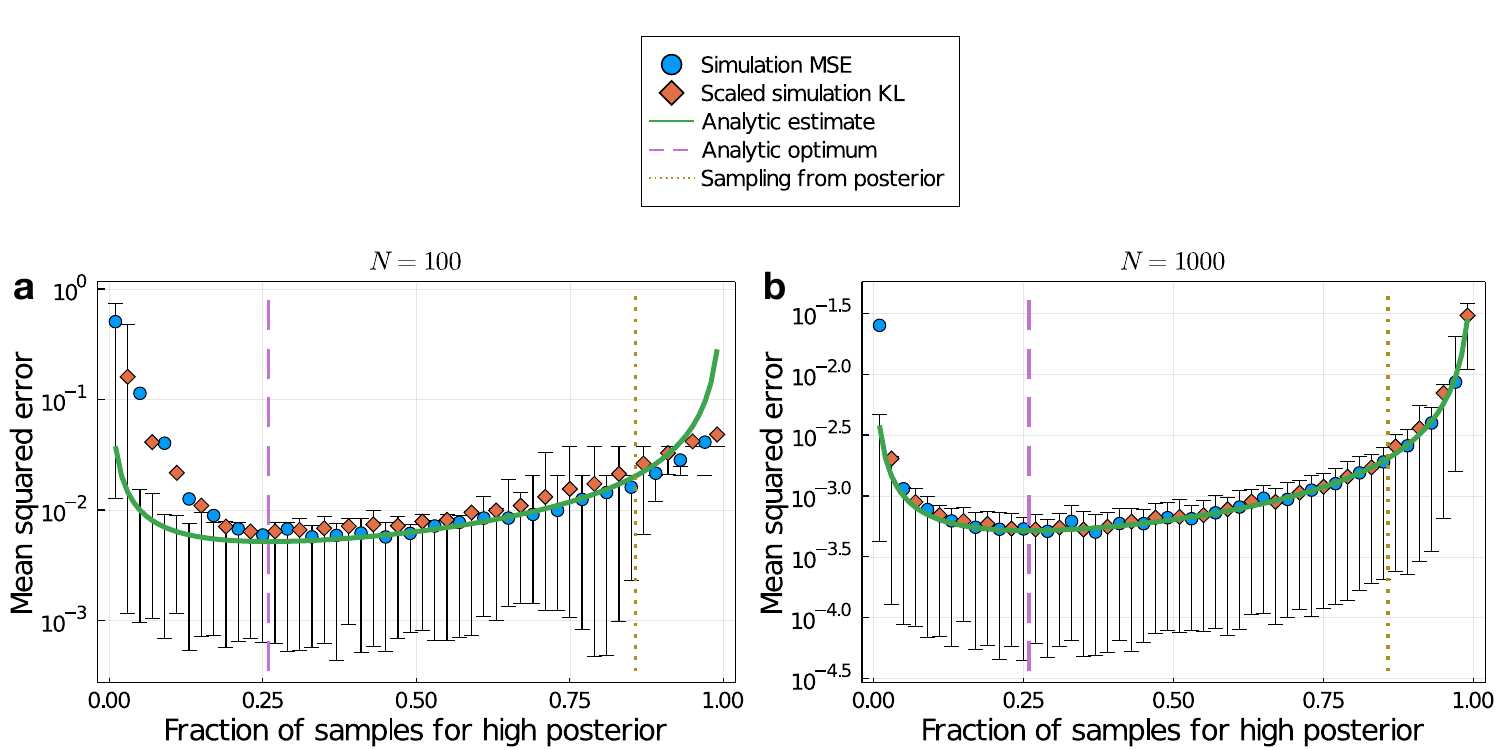}
\caption{The variance estimate in Eq.~\ref{e:discrete_asymptotic_variance} is qualitatively correct for small samples and precise for large samples, based on simulations  of a model selection task with $k=2$, $p(x^*|\theta_1) = 0.3$, and $p(x^*|\theta_2) = 0.05$. Each plot compares the mean squared error for the indicator function $f(\theta) = \mathbbm{1}(\theta = \theta_1)$ (blue dots) and mean KL divergence $KL\left(\hat p(\theta|x^*)| p(\theta|x^*)\right)$ scaled following Eq.~\ref{e:phi_divergence_approximation} (red diamonds), both averaged over 1000 trials, to Eq.~\ref{e:discrete_asymptotic_variance} (green line). The horizontal axis is the single algorithm parameter $n_1/N$. The optimum according to Eq.~\ref{e:discrete_optimal_n} (pink dashed line) is far from the posterior $n_1/N = p(\theta_1|x^*)$ (brown dotted line).
Error bars span the 75th and 25th percentiles over trials, which do not always cover the mean.
On the far right of (a), often no samples from $\theta_2$ are accepted, so $\hat p(\theta_1|x^*) = 1$; the error bar quantiles are the KL divergence or MSE for that estimate.
}
\label{f:two_sample_MSE_simulations}
\end{figure}

Figure~\ref{f:two_sample_MSE_simulations} illustrates the accuracy of the delta method calculation for an example with $p(x^*|\theta_1) = 0.3$, and $p(x^*|\theta_2) = 0.05$. 
With $N = 100$ simulations, the variance approximation in Eq.~\ref{e:discrete_asymptotic_variance} (solid green line) has the same scale and qualitative shape as empirical results, with some quantitative differences particularly for $n_1/N < 0.2$ where the expected number of samples with $x_i = x^*$ is less than 10.
There is a small visible difference between the rescaled average KL divergence $\mathbb{E}\left[KL\left(\hat p(\theta|x^*)|p(\theta|x^*)\right)\right]$ and the MSE of $\mathbb{E}[\mathbbm{1}(\theta = \theta_1)] = p(\theta_1|x^*)$.
As expected, our asymptotic results are more accurate for higher $N$.
With $N= 1000$, both bias and higher order contributions to the variance are negligible and the approximation of the KL divergence from Eq.~\ref{e:phi_divergence_approximation} is excellent.

The behavior for $k=2$ is not generic. For $k>2$, Equation~\ref{e:discrete_optimal_n} depends on $f$, as we illustrate for a larger discrete space and several target functions in the supplement (Section~\ref{app:discrete_gaussian}).
In addition, the factor $p_i^*$ penalizes parameters with low likelihoods. That penalization, however, can compete with the factor $|f(\theta_i) - \bar f|$ in typical situations where $\bar f$ is close to the value of $f(\theta)$ in regions with high posterior probability.
If, for example, $f(\theta) = \mathbbm{1}(\theta \in S)$, then $|f(\theta_i) - \bar f|$ is either $p(\theta \in S)$, if $\theta_i \notin S$, or $p(\theta \notin S)$, if $\theta_i \in S$.
The region with lower posterior probability should be sampled correspondingly more.

\subsection{Alternative scores offer alternative recommendations}

\begin{figure}
\center
\includegraphics[width=0.8\linewidth]{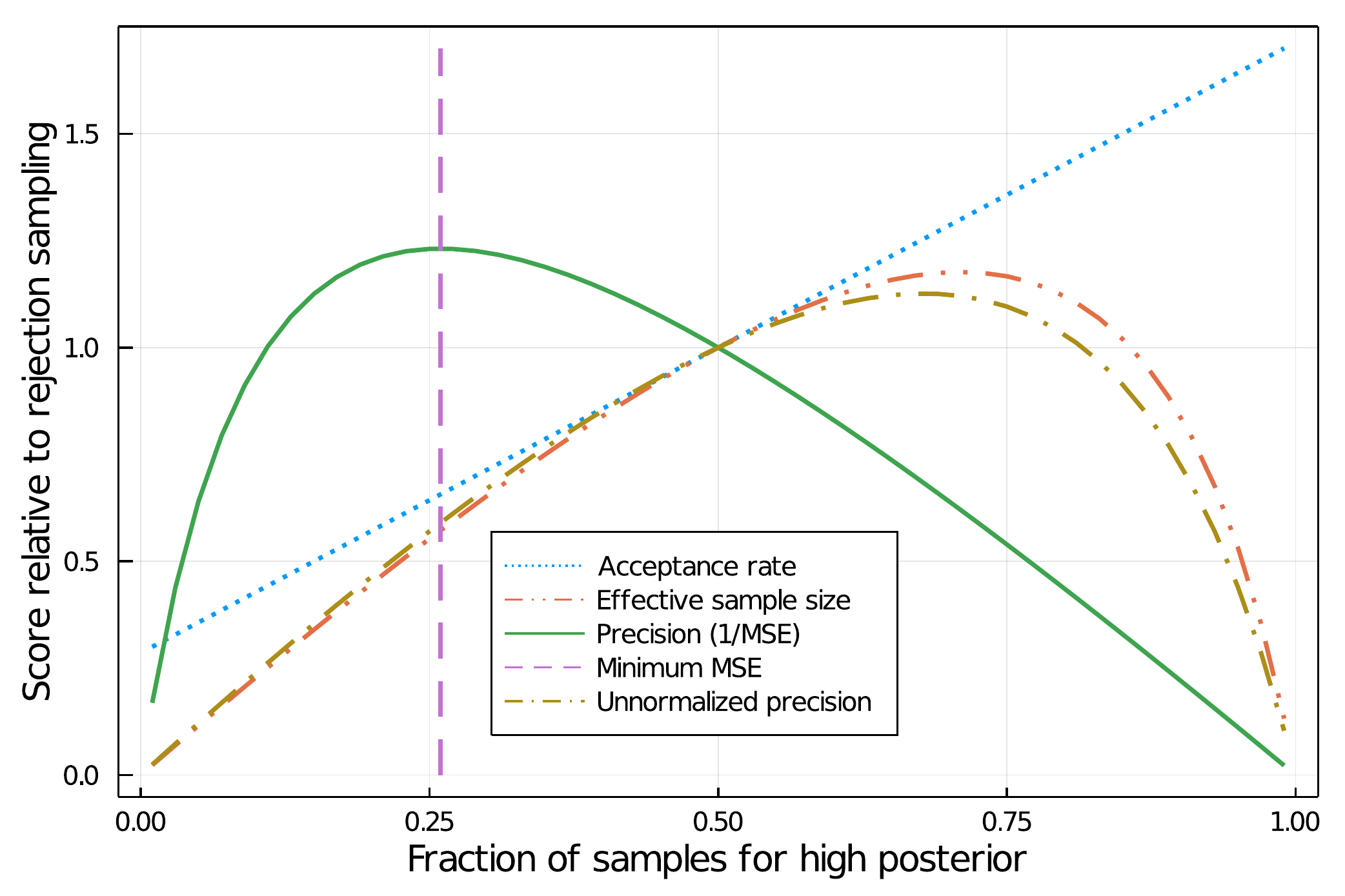}
\caption{
In the example of Fig.~\ref{f:two_sample_MSE_simulations}, the acceptance rate (blue, dotted), effective sample size (red, dash and two dots), or precision in estimating the unnormalized posterior (brown, dot-dashed) 
poorly match the precision in estimating the normalized posterior (green, solid).
The horizontal axis is the proportion of simulations performed with $\theta_1$, $n_1/N$.
Each curve shows the corresponding score divided by the same score for rejection sampling ($n_1/N = 0.5$).
}
\label{f:two_sample_score_comparison}
\end{figure}

The choice of score is critical in the derivation of the optimal sampling proportions in Eq.~\ref{e:discrete_optimal_n}.
Both acceptance rate and effective sample size are always improved by sampling more from parameters with higher likelihood, while error in the unnormalized posterior is minimized by sampling more from the parameter with likelihood closest to $0.5$ where the Bernoulli variance $p_i^*(1-p_i^*)$ is greater.
In the opposite direction, inverse binomial sampling~\cite{VanOpheusden2020}, where $n_i \propto 1/p_i^*$ asymptotically, samples from parameters in reverse order of likelihood.

We illustrate the different scores for the same two-hypothesis example as Fig.~\ref{f:two_sample_MSE_simulations} in Fig.~\ref{f:two_sample_score_comparison} and summarize their recommendations in Table~\ref{tab:optimal_ni}.
For ease of comparison to the acceptance rate and effective sample size, where higher is better, we plot the inverse MSE, or precision, rather than the MSE itself.
Optimizing $n_i$ for the effective sample size (Supplement Section~\ref{app:optimal_ESS_distribution}), for example, increases the expected $\widehat{ESS}$ by a factor of 1.18 relative to rejection sampling while decreasing the accuracy of the posterior estimate by a factor of 0.62.
IBS is the only strategy for choosing simulation parameters we found in the literature that outperforms rejection sampling in this example.
In general discrete spaces with $k>2$,  targeting an alternate score may or may not improve on rejection sampling in MSE depending on the likelihood and function of interest (Supplement Section~\ref{app:discrete_gaussian}).

\begin{table}
\begin{center}
\begin{tabularx}{1\linewidth}{p{0.3\linewidth} p{0.45\linewidth} X}
Method & Asymptotic optimal discrete simulation distribution & Efficiency relative to rejection sampling in Fig.~\ref{f:two_sample_score_comparison}\\
\hline
Maximize acceptance rate & $n_i \propto \mathbbm{1}(\theta_i \in \argmax_\theta p(x^*|\theta))$& 0\\
Inverse binomial sampling & $n_i \propto 1/p(x^*|\theta_i)$ & 1.11\\
Maximize $\widehat{ESS}$ &  $n_i \propto p(\theta_i) \sqrt{p(x^*|\theta_i)}$ & 0.62\\
Minimize unnormalized MSE & $n_i \propto p(\theta_i)  \sqrt{p(x^*|\theta_i)(1 - p(x^*|\theta_i))}$& 0.68\\
Minimize expectation MSE &$n_i \propto p(\theta_i)  \sqrt{p(x^*|\theta_i)(1 - p(x^*|\theta_i))}|f(\theta_i) - \bar f|$ & 1.23\\
\end{tabularx}
\end{center}
\caption{Distributions of simulations that optimize various scores. Maximizing the acceptance rate implies never simulating with some parameters, which does not give a consistent estimator.}
\label{tab:optimal_ni} 
\end{table}

When $f(\theta)$ is chosen appropriately, Eq.~\ref{e:discrete_optimal_n} optimizes other scores in addition to the MSE, in line with the derivations in Section~\ref{s:score_descriptions}. First, if $f(\theta)$ is an indicator function corresponding to a $50\%$ credible interval, so that $|f(\theta) - \bar f| = 0.5$ everywhere, we have $n_i \propto \pi_i \sqrt{p_i^* (1-p_i^*)}$, which optimizes the error in the unnormalized posterior (Supplement Section~\ref{app:optimal_simulation_distributions}). If we also take the limit $p(x^*|\theta) \rightarrow 0$ where $(1 - p(x^*|\theta))$ is constant, we are left with $n_i \propto \pi_i \sqrt{p_i^*}$ which is the proposal distribution that maximizes $\widehat{ESS}$~(\cite{Fearnhead2012} and Supplement Section~\ref{app:optimal_simulation_distributions}).

In that limit, the variance approximation in Eq.~\ref{e:discrete_asymptotic_variance} with simulations distributed according to either prior or posterior is $(4Np(x^*))^{-1}$ (Supplement Section~\ref{app:independent_sampling}).
Both are inefficient, as argued in a slightly different setting and way in~\cite{Li2018}.
For the prior, $(4Np(x^*))^{-1} \approx(4N_{acc})^{-1}$.
The acceptance rate does appear as a relevant score: a low acceptance rate suggest higher variance for a given dataset.
This justifies the intuition behind the standard argument against rejection sampling.
However, there is an important distinction between ``for a fixed sampling scheme, does a higher acceptance rate for dataset A over dataset B mean lower error for dataset A?'' and ``for a fixed dataset, does a higher acceptance rate for algorithm X over algorithm Y mean lower error for algorithm X?''.
Low acceptance rates indicate a difficult problem but not necessarily an inefficient algorithm.

Optimal performance by the MSE score is not immediately available, because Eq.~\ref{e:discrete_optimal_n} depends on the unknown likelihood and posterior mean.
 Given a target function $f(\theta)$, however, the optimal distribution of simulation parameters is straightforward to estimate adaptively.
Following that distribution, we can compute both an estimate of $\bar f$ with minimal asymptotic variance and an estimate of the variance itself (Supplement Section~\ref{app:discrete_adaptive}).
For brevity, we postpone that discussion to the supplement and end our treatment of the discrete case here.
We next turn to continuous parameter spaces, where despite new analytical challenges we will see qualitatively similar behavior.

\section{Continuous parameters}
\label{s:continuous_parameters}

Analysis like that of Section~\ref{s:discrete_parameters} is significantly more complicated when the parameter space $\Theta$ is continuous, for three main reasons.
\begin{enumerate}
\item It is no longer possible to simulate $n_i\rightarrow\infty$ times from each parameter value $\theta_i$. Instead, we must either deterministically choose or randomly sample $\{\theta_i\} \subset \Theta$. If using an importance-weighted average of accepted samples (Eq.~\ref{e:rejection_sampling_expectation}) the latter adds variance from the selection of $\{\theta_i\}$, which we calculate and optimize in the supplement (Section~\ref{app:sampling_parameters}), while the former option introduces bias.
\item There are more viable strategies for estimating $\mathbb{E}_{p(\theta|x^*)}[f(\theta)]$ given the simulation results.
One could use Eq.~\ref{e:rejection_sampling_expectation}, apply multilevel Monte Carlo for variance reduction~\cite{Warne2018}, or numerically integrate an approximate posterior based on a kernel regression estimate~\cite{Biau2015, Blum2010}, Gaussian process~\cite{Gutmann2016,Jarvenpaa2019}, or mixture density network~\cite{ Lueckmann2017, Papamakarios2016}.
As we will see, this choice can have a significant effect on the quality of the estimate.
\item It may be difficult to choose or sample simulation parameters from the optimal distribution. In the discrete case, the minimum-MSE $n_i$ from Eq.~\ref{e:discrete_optimal_n} can be explicitly computed and normalized given a posterior approximation.
In the continuous case, sampling likely requires either a simplification of the optimal distribution or MCMC with additional technical considerations and computational overhead.
\end{enumerate}

We do not fully address these obstacles.
Instead, we present three examples demonstrating that results for the discrete case qualitatively translate.
First, we show that for estimating the posterior mean of a simple one-dimensional parameter the optimal distribution for independently sampled $\theta$ outperforms sampling from the prior or posterior or maximizing $\widehat{ESS}$.
Second, as an illustration of the additional complexity of continuous parameters, we present an example where changing from Eq.~\ref{e:rejection_sampling_expectation} to numerically integrating a kernel regression estimate (Supplement Section~\ref{app:kernel_regression}) reduces MSE similarly to optimizing the choice of simulation parameters.
Finally, we apply the optimized independent sampling distribution in a model selection problem combining discrete and continuous parameters.

\begin{figure}
\includegraphics[width= \linewidth]{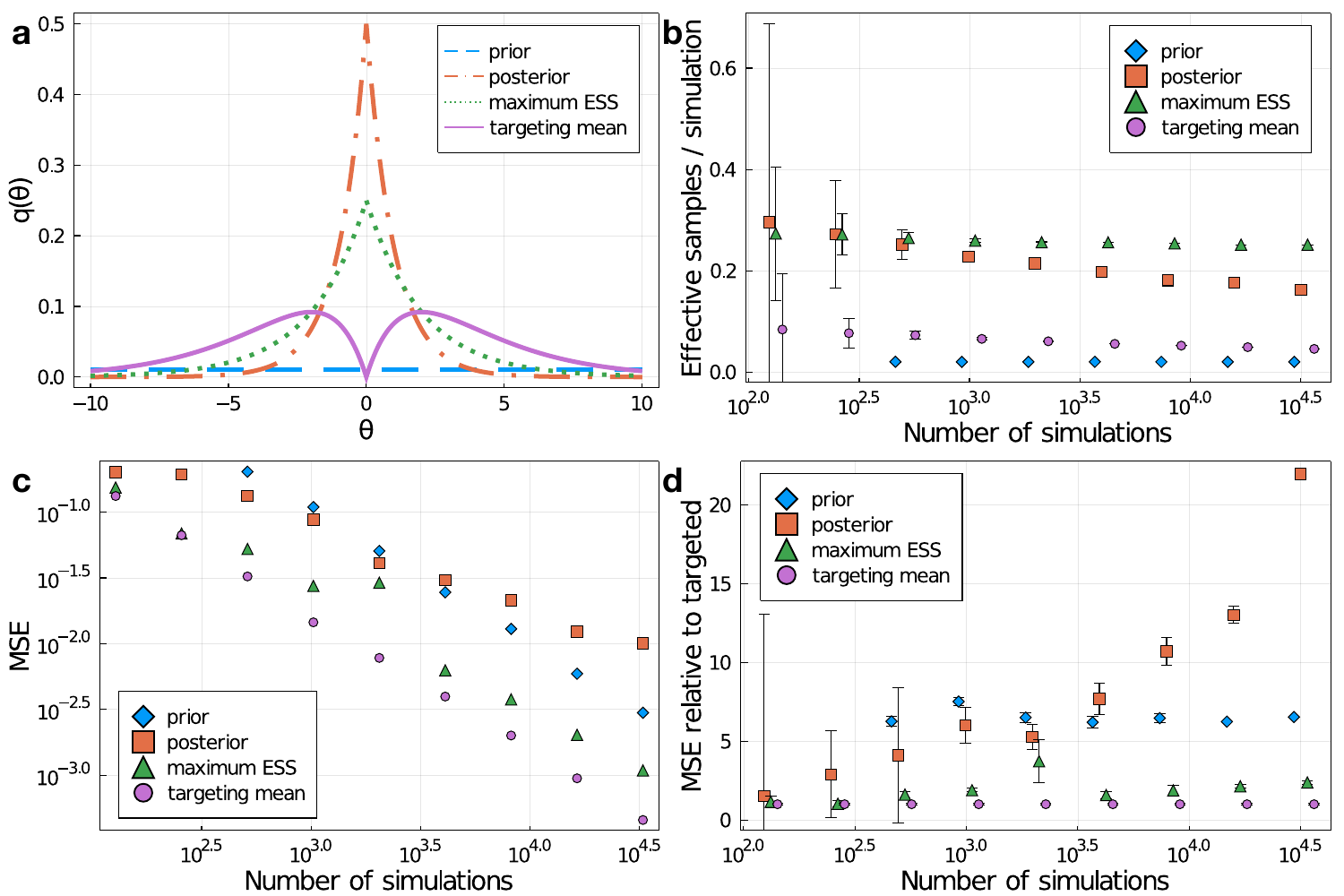}
\caption{Sampling from a distribution targeting $f(\theta) = \theta$ yields lower error than generic strategies.
Here we consider a one dimensional parameter $\theta \in \mathbb{R}$ with prior uniform on $[-40, 60]$ and likelihood $p(x^*|\theta) = e^{-|\theta|}$.
(a) Our candidate sampling distributions are the prior (blue, dashed), posterior (red, dot-dashed), $q_{ESS}(\theta)\propto p(\theta) \sqrt{p(x^*|\theta)}$ to maximize $\widehat{ESS}$ (green, dotted), and $q_{f}(\theta)\propto (f(\theta) - \bar f) p(\theta) \sqrt{p(x^*|\theta)}$ (purple, solid) to optimize the MSE with independent sampling (Supplement Section~\ref{app:independent_sampling}).
We ran 1000 trials and computed the empirical mean effective sample size (b) and squared error in estimating $\mathbb{E}_{p(\theta|x^*)}[\theta]\approx 0$, shown on an absolute scale in (c) and relative to the error when using $q_f(\theta)$ in (d).
The targeted distribution (purple circles) has an effective sample size  $\sim 3\times$ higher than sampling from the prior (blue diamonds) but $ \sim 4\times$ lower than the maximum (green triangles).
Despite the lower $\widehat{ESS}$, the MSE for the targeted distribution is $\sim 2\times$ lower than for the $\widehat{ESS}$-optimizing distribution.
The observed performance of posterior sampling (red squares) by either metric gets relatively worse with larger $N$, likely because more of our trials include accepted simulations with low likelihood and high weight that decrease $\widehat{ESS}$ and increase variance.
Points are omitted for prior sampling with $N<200$ because there is a significant chance no simulations are accepted and no posterior estimate can be made.
Error bars in (b) show empirical standard errors. Error bars in (d)  show empirical standard errors ignoring variance of the denominator; judging by the variation across $N$, they likely underestimate variability.
The horizontal position of points in (b) and (d) is shifted slightly so the error bars do not overlap.
}
\label{f:continuous_simple}
\end{figure}

For independently sampled parameters, a delta method calculation similar to the discrete case (Supplement Section~\ref{app:sampling_parameters}) leads to an optimal importance distribution
\begin{equation}
q_{f,ind}(\theta) \propto p(\theta) \sqrt{p(x^*|\theta)}|f(\theta) - \bar f|.
\label{e:independent_sampling_optimal_n}
\end{equation}
In our first example (Fig.~\ref{f:continuous_simple}), we let $\theta \in \mathbb{R}$ with prior uniform on $[-40, 60]$ and likelihood $e^{-|\theta|}$. 
We evaluate the mean squared error in $\mathbb{E}_{p(\theta|x^*)}[\theta]$ with four sampling strategies: prior sampling $\theta_i \sim p(\theta)$, posterior sampling $\theta_i \sim p(\theta|x^*)$, optimization of $\widehat{ESS}$ with $\theta_i \sim q_{ESS}(\theta) \propto p(\theta) \sqrt{p(x^*|\theta)}$, and optimized independent sampling with $\theta_i \sim q_{f,ind} (\theta)$.

Sampling from the prior requires at least $p(x^*)^{-1} \approx 100$ samples to consistently have nonzero acceptances; beyond that threshold, it performs as well as or better than sampling from the posterior (Fig.~\ref{f:continuous_simple}). Optimizing for the effective sample size is better than either, and including the factor $|f(\theta) - \bar f| = |\theta|$ improves accuracy (Fig.~\ref{f:continuous_simple}c,d) while decreasing $\widehat{ESS}$ (Fig.~\ref{f:continuous_simple}b).

\label{s:KDE}

Although the targeted distribution outperformed alternatives in Fig.~\ref{f:continuous_simple}, there are two possibilities for improvement. 
First, estimating $\bar f$ with the empirical average in Eq.~\ref{e:rejection_sampling_expectation} need not be optimal.
Instead, we could train a model to explicitly compute the likelihood or posterior and integrate the result. 
Much of the recent research effort in likelihood-free inference has been devoted to developing methods to do that.

One approach models the distribution of the data $x$ or discrepancy $\Delta(x, x^*)$ as a function of $\theta$.
For example, \textit{Bayesian synthetic likelihood}~\cite{Price2018}
fits a multivariate normal distribution to observed summary statistics for each $\theta$ considered,
while J\"arvenp\"a\"a et al.~\cite{Jarvenpaa2019} use a Gaussian process prior
on $\Delta$.
The resulting model then gives an estimate of the likelihood of the observed data $p(x^*|\theta)$.

Other proposals~\cite{Alsing2019, Durkan2018, Fengler2021, Lueckmann2017, Lueckmann2018, Papamakarios2019}
use density estimators based on neural networks to approximate either $p(x^*|\theta)$ (as a function of $\theta$) or $p(x|\theta)$ (as a function of both $x$ and $\theta$).
The likelihood estimate is then combined with the prior to form a posterior, which can also be targeted directly~\cite{Papamakarios2016}.
Alternatively, kernel regression estimates (\cite{Biau2015,Blum2010}, Supplement Section~\ref{app:kernel_regression})  convert an empirical distribution of samples to a continuous posterior estimate.
For ABC-SMC, accuracy can be improved by including accepted samples from all rounds in the final output (Supplement Section~\ref{app:ABC-SMC_keeping_all_rounds}).
Each of these algorithms is a proposal for step 2 of likelihood-free inference and can be combined with any strategy for choosing the simulation parameters $\{\theta_i\}$.

The second way of improving estimates is to sample $\{\theta_i\}$ with variance reduction, for example via stratification (Supplement Section~\ref{app:stratified}). This can reduce the part of the variance in Eq.~\ref{e:rejection_sampling_expectation} that comes purely from sampling parameters.
Thoroughly investigating either of these choices goes beyond the scope of this paper. Instead, we show one example where both stratification and changing from estimating as the empirical weighted mean of accepted samples (Eq.~\ref{e:rejection_sampling_expectation}) to numerical integration of a kernel regression estimate give similar improvements in accuracy.

\begin{figure}
\includegraphics[width=\linewidth]{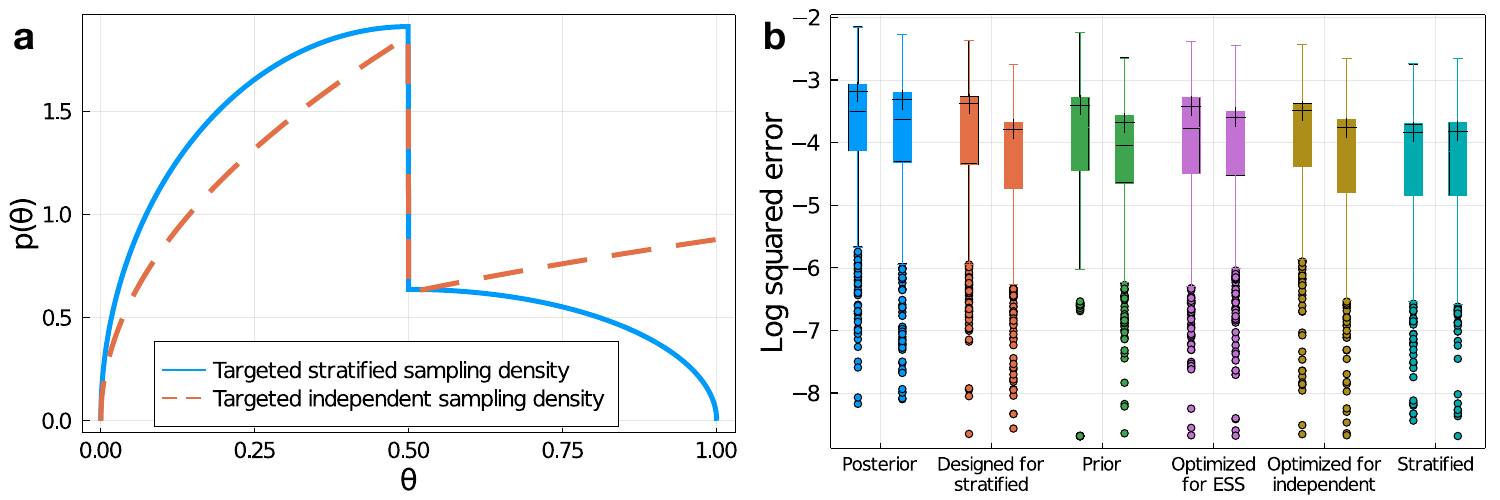}
\caption{
The optimal distribution of simulation parameters depends on whether parameters are sampled independently and how posterior expectations are estimated.
Here we consider estimating the posterior probability that $\theta < 0.5$ (via $f(\theta) = \mathbbm{1}[\theta < 0.5]$) with $\theta \in [0,1]$ and $p(x^*|\theta) = \theta$.
Panel (a) shows distributions targeted at $f(\theta)$  designed for stratified (solid blue line, Eq.~\ref{e:stratified_optimal_n}) or independent (dashed red line, Eq.~\ref{e:independent_sampling_optimal_n}) sampling of $\{\theta_i\}$.
In (b), we compare the distribution of $\log_{10}\left((\mathbb{E}_{\hat p}[f(\theta)] -\bar f)^2\right)$
for stratified sampling using Eq.~\ref{e:stratified_optimal_n} (cyan) to five different independent sampling strategies: (blue) sampling from the posterior, (red) sampling independently from the distribution designed for stratified sampling, (green) sampling from the prior, (purple) sampling from $p(\theta) \sqrt{p(x^*|\theta)}$ to maximize $\widehat{ESS}$, and (brown) sampling from the optimal independent sampling distribution (Eq.~\ref{e:independent_sampling_optimal_n}).
For each sampling distribution, the left box plot shows the error using Eq.~\ref{e:rejection_sampling_expectation} while the right box plot shows the error from numerically integrating a kernel regression estimate (Supplement Section~\ref{app:kernel_regression}), both with $N=1000$ simulations.
Crosses mark the mean of each distribution.
While stratified sampling following Eq.~\ref{e:stratified_optimal_n} performs the best, other sampling strategies have similar error distributions when using the kernel regression estimate.
}
\label{f:KDE}
\end{figure}

Fig.~\ref{f:KDE} presents a case where optimizing for $\widehat{ESS}$ performs no better than rejection sampling.
The parameter $\theta$ has a uniform prior on $[0,1]$, $p(x^*|\theta) = \theta$, and we are interested in the posterior probability that $\theta < 0.5$.
Stratified sampling following the continuous equivalent to Eq.~\ref{e:discrete_optimal_n} (Supplement Section~\ref{app:stratified}) performs better than any other sampling strategy, but alternative distributions have nearly the same MSE if a kernel regression estimate is used.
Full details of our kernel approach are given in the supplement (Section~\ref{app:kernel_regression}).
This both is consistent with the good results seen with recent methods~\cite{Alsing2019, Durkan2018, Fengler2021,Jarvenpaa2019, Lueckmann2017, Lueckmann2018, Warne2018} that change how the posterior is estimated from simulations and suggests there is room for improvement in selection of simulation parameters.

\begin{figure}
\includegraphics[width= \linewidth]{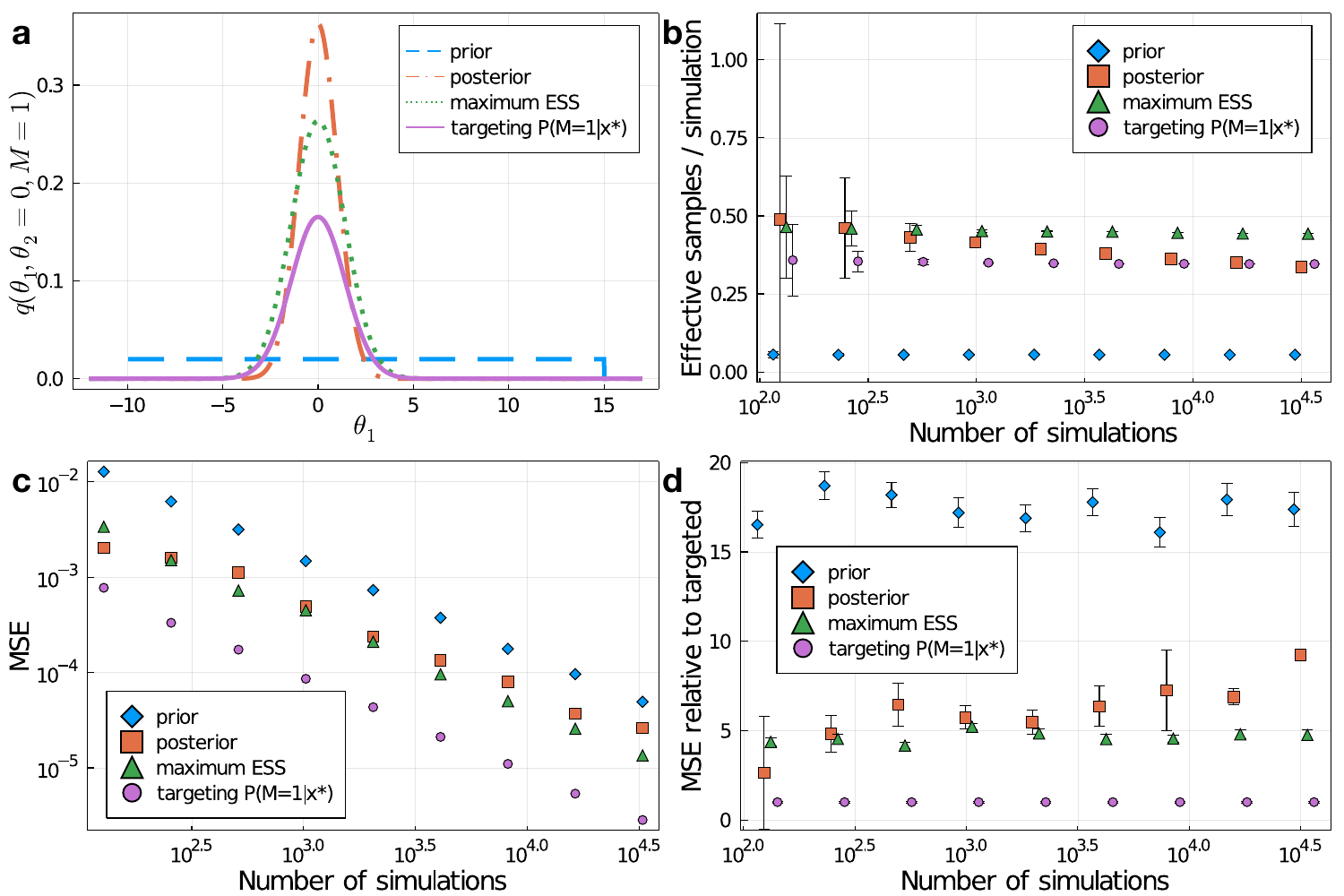}
\caption{
Targeting the posterior probability of Model 1 improves accuracy.
(a) Our candidate sampling distributions are the prior (blue, dashed), posterior (red, dot-dashed), $q_{ESS}(\theta)\propto p(\theta) \sqrt{p(x^*|\theta)}$ to maximize $\widehat{ESS}$ (green, dotted), and $q_f(\theta)\propto (f(\theta) - \bar f) p(\theta) \sqrt{p(x^*|\theta)}$ (purple, solid) with $f(\theta) = \mathbbm{1}[M = 1]$ to optimize the MSE with independent sampling (Supplement Section~\ref{app:independent_sampling}).
For each, we plot the importance distribution evaluated at $M=1$, $\theta_2 =0$, as a function of $\theta_1$.
We ran 1000 trials and computed the empirical mean effective sample size (b) and squared error in estimating $p(M = 1|x^*)$, shown on an absolute scale in (c) and relative to the error when using $q_f(\theta)$ in (d).
The targeted distribution (purple circles) has an intermediate effective sample size $ \sim 20\%$ lower than the optimum (green triangles) but substantially higher than sampling from the prior (blue diamonds).
Despite the lower $\widehat{ESS}$, the MSE for the targeted distribution is $\sim 5\times$ lower than for the $\widehat{ESS}$-optimizing distribution.
The observed performance of posterior sampling (red squares) by either metric gets relatively worse with larger $N$, likely because more of our trials include accepted simulations with low likelihood and high weight that decrease $\widehat{ESS}$ and increase variance.
Error bars in (b) show empirical standard errors. Error bars in (d)  show empirical standard errors ignoring variance of the denominator; judging by the variation across $N$, they likely underestimate variability.
The horizontal position of points in (b) and (d) is shifted slightly so the error bars do not overlap.
}
\label{f:discrete+continuous}
\end{figure}

Our final example is a combination of the discrete and continuous cases.
This represents a model selection problem~\cite{Toni2009, Toni2009a}, where there is at least one discrete parameter $M$ (representing the choice of model) as well as model-specific parameters that may be continuous. 
Our results can be straightforwardly applied to such cases.
As an illustration, we consider a problem with two models of equal prior probability.
The likelihood for model 1  ($M=1$) is $e^{-\theta_1^2/2}$; the likelihood for model 2 ($M = 2$) is $e^{-(\theta_1^2 + \theta_2^2)/2}$. For each model, the prior on each $\theta_i$ is uniform between $-10$ and $15$.
Without a subscript, $\theta = (M, \theta_1, \theta_2)$ refers to all three parameters together.
As in Fig.~\ref{f:continuous_simple}, we compare four sampling strategies:  $\theta_i \sim p(\theta)$,  $\theta_i \sim p(\theta|x^*)$,  $\theta_i \sim q_{ESS}(\theta) \propto p(\theta) \sqrt{p(x^*|\theta)}$, and $\theta_i \sim q_{f,ind} (\theta) \propto  |f(\theta)-\bar f|p(\theta) \sqrt{p(x^*|\theta)}$.
We choose $f(\theta) = \mathbbm{1}[M = 1]$ to target the posterior probability of model 1.

The evidence for model 1 is high: model 2 has an extra parameter and more diffuse prior without a better maximum likelihood. For the true posterior $\mathbb{E}_{p(\theta|x^*)}[f(\theta)] \approx 0.9$.
The proportion of simulations assigned to Model 1 is 93\% for maximizing $\widehat{ESS}$ but only 58\% for targeted independent sampling (Eq.~\ref{e:independent_sampling_optimal_n}).
Again, incorporating the target function $f$ yields higher precision (Fig.~\ref{f:discrete+continuous}c-d) and lower effective sample size (Fig.~\ref{f:discrete+continuous}b).

\section{Discussion}
\label{s:discussion}

Properly evaluating the many important recent advances in likelihood-free inference requires clear quantitative scores for algorithm performance.
We argue for scoring by the mean squared error of estimated function expectations, on the grounds that that score can be usefully applied to any algorithm yielding an approximate posterior, captures key features of interest like the mean, variance, and credible intervals, and includes the most promising other alternatives as limiting special cases.

This MSE score explicitly depends on the target $f(\theta)$, which we see as an advantage.
Common alternatives, like the effective sample size $\widehat{ESS}$, leave that dependence implicit and possibly unacknowledged.
Other scores answer different questions: the acceptance rate, for example, is a better measure of the difficulty of a problem than the efficiency of an algorithm.
Our empirical results show that using distributions targeted at specific functions, preferably with stratified rather than independent sampling, leads to lower MSE than optimization for generic scores.

Past work~\cite{Barber2015,Blum2010} has analyzed the MSE score for rejection sampling. 
Papers introducing more sophisticated algorithms, however, have often used heuristics like the acceptance rate or effective sample size for evaluation~\cite{Bonassi2015,DelMoral2012,Filippi2013,Meeds2015, Prescott2021,Price2018, Sisson2007}.
We are not aware, for example, of a treatment of the variance of ABC-SMC estimators comparable to Barber et al's analysis of ABC rejection~\cite{Barber2015}.
This paper begins to bridge that gap by optimizing the distribution of simulation parameters directly for the MSE score.

Carefully measuring accuracy allows us to disprove the claim~\cite{Sisson2007} that sampling simulation parameters from the posterior, as is targeted in a wide range of algorithms~\cite{Beaumont2009, DelMoral2012, Filippi2013,Lueckmann2017, Papamakarios2016,Papamakarios2019,Toni2009a, Toni2009}, is optimal for ABC, or even consistently better than sampling from the prior.
This complements earlier work arguing against sampling from either prior or posterior by comparing to maximum likelihood (ML) estimation in an informative-data limit where ML works well~\cite{Li2018}.
The intuition underlying Eq.~\ref{e:discrete_optimal_n}, that simulations should be concentrated where (1) they are required to learn the likelihood and (2) learning the likelihood matters, should be generally relevant for all state-of-the-art methods.

Our results provide a base for further research in many directions. First, all of our analysis is in the limit of large sample sizes. With a small number of simulations, higher order terms ignored with the delta method as well as the difference between a $\phi$-divergence and expectation MSE may be significant.
 Future work may also consider the effect of the supremum over a class of functions in an integral probability metric.
MMD~\cite{Gretton2012} in particular is a promising candidate for further analysis, as unlike TV and Wasserstein it converges with rate $n^{-1/2}$ for empirical samples.

A rigorous treatment of the case with continuous parameters would be worthwhile, as would more analysis of continuous or high-dimensional data where summary statistics are commonly used~\cite{Fearnhead2012}.
There are open questions both in how to choose and sample from an optimized distribution of parameters and in how to estimate the likelihood and posterior given simulation results; as we see in Fig.~\ref{f:KDE}, each step matters.

Following much of the ABC literature~\cite{Barber2015,Prescott2021,Warne2018}, we assume that the primary computational cost of likelihood-free inference is in simulating from the model and ignore algorithmic overhead. 
That assumption could be investigated more carefully; one particular risk is that an overly complex algorithm may lose the easy parallelization available with rejection sampling.
For sufficiently large $N$ and cheap simulations, it will also be important to choose an algorithm with $O(N)$ complexity~\cite{DelMoral2012} rather than the $O(N^2)$ of naive ABC-SMC.

Finally, we assumed the goal was to learn the posterior for a single observed sample $x^*$.
In some applications, the same simulations may be used to analyze multiple datasets separately~\cite{Fengler2021}.
Optimal simulation distributions for such \textit{global learning}~\cite{Alsing2019,Lueckmann2018} are not yet known, though sampling from the prior~\cite{Papamakarios2016} is intuitively plausible.

For all of these research directions, the choice of performance score will continue to matter.
Future work should pay careful attention to appropriately measuring accuracy and efficiency.

\section*{Code availability}
Julia code to reproduce our examples is available at \url{https://github.com/aforr/LFI_accuracy}.
\bibliographystyle{plain}
\bibliography{ABC_references.bib}

\title{Measuring the accuracy of likelihood-free inference\\Supplementary material}
\emptythanks
\maketitle
\setcounter{section}{0}
\renewcommand{\thesection}{S\arabic{section}}
\setcounter{table}{0}
\renewcommand{\thetable}{S\arabic{table}}
\setcounter{figure}{0}
\renewcommand{\thefigure}{S\arabic{figure}}
\setcounter{equation}{0}
\renewcommand{\theequation}{S\arabic{equation}}

\section{Reverse $\phi$-divergence approximation}
\label{app:phi_divergence}
Here we show that in the limit where a $\phi$-divergence with smooth $\phi$ is small in both directions, Eq.~\ref{e:phi_divergence_approximation} gives an approximation to $d_{\phi\text{-div}}(q,\hat q)$ as well as $d_{\phi\text{-div}}(\hat q,q)$.
Following the derivation in the main text, a second order Taylor expansion of $\phi\left(q(\theta)/\hat q(\theta)\right)$ leads to 
\begin{equation}
d_{\phi\text{-div}}(q,\hat q) =\int \left[\frac{\phi''(1)}{2} \frac{\left(\hat q(\theta) - q(\theta)\right)^2}{\hat q(\theta)} +  O\left(\left(\frac{\hat q(\theta)}{q(\theta)} - 1\right)^3\right)\right]\dd\theta.
\end{equation}
We can then substitute 
\begin{equation}
\hat q(\theta) ^{-1}= q(\theta)^{-1} \left(1 + \left(\frac{\hat q(\theta)}{q(\theta)} - 1\right)\right)^{-1} =q(\theta)^{-1} \left(1 - \left(\frac{\hat q(\theta)}{q(\theta)} - 1\right)\right) + O\left(\left(\frac{\hat q(\theta)}{q(\theta)} - 1\right)^2\right),
\end{equation}
to find
\begin{equation}
d_{\phi\text{-div}}(q,\hat q) =\int \left[\frac{\phi''(1)}{2} \frac{\left(\hat q(\theta) - q(\theta)\right)^2}{q(\theta)} +  O\left(\left(\frac{\hat q(\theta)}{q(\theta)} - 1\right)^3\right)\right]\dd\theta.
\end{equation}
Like for the $\phi$-divergence in the opposite direction, the final leading order term is equal to the expectation mean squared error $d_f(\hat q, q)$ with target function
 \begin{equation}
 f(\theta) = \left(\frac{\phi''(1)}{ 2q(\theta)}\right)^{1/2}\delta(\theta' - \theta).
 \end{equation}
While $\phi$-divergences are in general asymmetric, the asymmetry shows up only at higher order in $\hat q(\theta)/q(\theta) - 1$.

\section{Delta method for ratio of random variables}
\label{app:delta_method}
Many of the estimators in this paper are of the form
\begin{equation}
\mathbb{E}_{\hat p(\theta|x^*)}\left[ f(\theta) \right] = \frac{R}{S},
\end{equation}
where $R$ and $S$ are random variables with individual means $\mu_R$ and $\mu_S$, respectively, such that $\mu_R/\mu_S = \bar f$ is the true posterior mean.
We would like to estimate the mean squared error
\begin{equation}
\mathbb{E}\left[\left(\frac{R}{S} - \frac{\mu_R}{\mu_S}\right)^2\right]
= \var\left(\frac{R}{S}\right) + \left(\mathbb{E}\left[ \frac{R}{S} \right] - \frac{\mu_R}{\mu_S} \right)^2.
\end{equation}

In the limit of infinite simulations $N \rightarrow \infty$, both $(R - \mu_R)^2$ and $(S - \mu_S)^2$ will be small, of order $1/N$.
We therefore perform a Taylor expansion of $g(R, S) = R/S$ about $g(\mu_R, \mu_S)$:
\begin{align}
g(R, S) & = 
g(\mu_R, \mu_S) + \partial_R g(\mu_R, \mu_S) (R - \mu_R) + \partial_S g(\mu_R, \mu_S) (S - \mu_S) + \Delta
\\
& = 
\frac{\mu_R}{\mu_S} + \frac{R - \mu_R}{\mu_S} - \frac{\mu_R(S - \mu_S)}{\mu_S^2}
+ \Delta
\end{align}
where the remainder is $\Delta =O\left((R - \mu_R)^2 +  (S - \mu_S)^2\right) = O( 1/N)$. 
For the bias term, the leading order terms all cancel leaving
\begin{align}
\left(\mathbb{E}\left[ \frac{R}{S} \right] - \frac{\mu_R}{\mu_S} \right)^2
= \left( \mathbb{E}[\Delta] \right)^2 = O(N^{-2}).
\end{align}
For the variance term, we drop all terms involving only $\mu_R$ and $\mu_S$, leaving
\begin{align}
\var\left(\frac{R}{S}\right)
& = \var\left(\frac{R}{\mu_S} - \frac{\mu_R S}{\mu_S^2} + \Delta\right) = \var\left(\frac{R}{\mu_S} - \frac{\mu_R S}{\mu_S^2}\right) + O(N^{-3/2}),
\end{align} 
where the last equality follows because $\Delta = O(N^{-1})$ and $R-\mu_R$ and $S - \mu_S$ are both $O(N^{-1/2})$.

Rearranging slightly,
\begin{align}
\var\left(\frac{R}{S}\right) &= \mu_S^{-2} \var \left(R - S\frac{\mu_R}{\mu_S}\right) + O(N^{-3/2})
\label{e:delta_method_reorganized}
\\
&=\frac{1}{\mu_S^2} \var(R) -2\frac{\mu_R}{\mu_S^3}\cov(R, S) + \frac{\mu_R^2}{\mu_S^4}\var(S).
\label{e:delta_method_appendix}
\end{align}

\section{Deriving optimal simulation distributions}
\label{app:optimal_simulation_distributions}

In this section, we derive the distributions $n_i$ over discrete parameters that optimize each of three scores: the asymptotic variance in the estimate of $\mathbb{E}_{p(\theta|x^*)}[f(\theta)]$; the integrated variance of the unnormalized posterior; and the effective sample size.
As in Section~\ref{s:discrete_parameters}, we consider taking $n_i$ samples from $\theta_i$, of which $n_i^*$ result in $x = x^*$. We then estimate the likelihood as $\hat p(x^*|\theta) = n_i^*/n_i$.

\subsection{Mean squared error in $\mathbb{E}_{p(\theta|x^*)}[f(\theta)]$}
\label{app:optimal_MSE_distribution}

The asymptotic variance we wish to minimize, from Eq.~\ref{e:discrete_asymptotic_variance}, is
\begin{align}
\var\left(
\mathbb{E}_{\hat p(\theta|x^*)} \left[ f(\theta) \right]
\right)
& =  \mu_S^{-2}\sum_{i=1}^k \pi_i^2 \frac{p_i^*(1-p_i^*)}{n_i}\left(
 f(\theta_i) - \bar f
\right)^2.
\end{align}
This has the form
\begin{align}
\var\left(
\mathbb{E}_{\hat p(\theta|x^*)} \left[ f(\theta) \right]
\right)
\approx\sum_{i=1}^k \frac{\alpha_i}{n_i},
\label{e:sum_of_reciprocals}
\end{align}
where we have condensed every factor independent of $n_i$ into the coefficient
\begin{equation}
\alpha_i = p(x^*)^{-2}p(\theta_i) p(x^*|\theta_i)(1 - p(x^*|\theta_i) \left(f(\theta_i) - \bar f\right)^2.
\label{e:MSE_alpha_i}
\end{equation}
The constrained minimum with respect to $n_i$ occurs at a stationary point of the Lagrangian
\begin{equation}
\mathcal{L} = \sum_{i=1}^k \frac{\alpha_i}{n_i} - \lambda\left(N - \sum_i n_i\right).
\end{equation}
This is a convex problem, so the unique local minimum is the global minimum. By setting derivatives with respect to $n_i$ to zero, we find
\begin{equation}
\frac{\alpha_i}{n_i^2} = \lambda,
\end{equation}
implying that
\begin{equation}
n_i^* = \sqrt{\frac{\alpha_i}{\lambda}} \propto \sqrt{\alpha_i}.
\label{e:sum_of_reciprocals_argmin}
\end{equation}
Substituting Eq.~\ref{e:MSE_alpha_i} into $n_i \propto \sqrt{\alpha_i}$  and dropping the constant factor $p(x^*)^{-1}$ yields Eq.~\ref{e:discrete_optimal_n}.

\subsection{Unnormalized posterior}
For a single parameter value $\theta_i$, $n_i^*$ has a binomial distribution with variance $n_ip(x^*|\theta_i)\left( 1 - p(x^*|\theta_i)\right)$.
The unnormalized posterior estimate $p(\theta_i) n_i^*/n_i$ then has variance $p(\theta_i)^2 p(x^*|\theta_i)\left( 1 - p(x^*|\theta_i)\right)/n_i$.
Summing over $\theta_i$,
\begin{equation}
\sum_{i=1}^k\var\left(
p(\theta_i) \hat p(x^*|\theta)
\right)
 = \sum_{i=1}^k \frac{p(\theta_i)^2 p(x^*|\theta_i)\left( 1 - p(x^*|\theta_i)\right)}{n_i}.
\label{e:unnormalized_variance}
\end{equation}
Eq.~\ref{e:unnormalized_variance} is of the same form as Eq.~\ref{e:sum_of_reciprocals} with $\alpha_i = p(\theta_i)^2 p(x^*|\theta_i)\left( 1 - p(x^*|\theta_i)\right)$; the corresponding constrained minimizer is 
\begin{equation}
n_i^* \propto p(\theta_i) \sqrt{p(x^*|\theta_i)\left( 1 - p(x^*|\theta_i)\right)}
\end{equation}
by Eq.~\ref{e:sum_of_reciprocals_argmin}.

\subsection{Effective sample size}
\label{app:optimal_ESS_distribution}
The distribution that optimizes $\widehat{ESS}$ for independently sampled continuous parameters is known in the literature~\cite{Fearnhead2012,Li2018}.
Here we derive the same result for our discrete case.

For each $i$, we have $n_i^*$ accepted samples each with weight $w_i = p(\theta_i)/n_i$.
The effective sample size is
\begin{equation}
\widehat{ESS}
=\frac{\left(\sum_{i=1}^k n_i^*w_i\right)^2}{\sum_{i=1}^k n_i^*w_i^2}
= 
\frac{\left(\sum_{i=1}^k \frac{n_i^*p(\theta_i)}{n_i}\right)^2}{\sum_{i=1}^k n_i^*\left(\frac{p(\theta_i)}{n_i}\right)^2}.
\end{equation}
 In the limit $n_i \rightarrow \infty$, $n_i^*/n_i$ converges in probability to $p(x^*|\theta)$. Substituting to eliminate $n_i^*$, we have
\begin{equation}
\widehat{ESS}
\overset{p}{\rightarrow}
\frac{\left(\sum_{i=1}^k p(x^*|\theta)p(\theta_i)\right)^2}{\sum_{i=1}^k \frac{p(x^*|\theta)p(\theta_i)^2}{n_i}}
= p(x^*)^2\left(
\sum_{i=1}^k \frac{p(x^*|\theta)p(\theta_i)^2}{n_i}
\right)^{-1}.
\end{equation}
Maximizing $\widehat{ESS}$ with respect to $n_i$ is equivalent to minimizing its reciprocal. Again using the Lagrange multiplier approach via Eq.~\ref{e:sum_of_reciprocals_argmin}, the maximizer is 
\begin{equation}
n_i^* \propto p(\theta_i) \sqrt{p(x^*|\theta)}.
\end{equation}

\section{Discrete parameter examples}
\label{app:discrete_examples}

\subsection{Discretized Gaussian}
\label{app:discrete_gaussian}
To illustrate our results for larger discrete parameter spaces than the $k=2$ case considered in the main text, here we investigate a discretized version of a Gaussian distribution.
The parameter space $\Theta$ consists of 101 evenly spaced points from $-5$ to $5$. 
The prior is uniform; the likelihood is $e^{-\theta^2/2}$.
Importantly, we treat the space as fully discrete. The likelihood for each $\theta_i$ is estimated independently; none of the algorithms are aware that similar values of $\theta$ yield similar likelihoods.
In a realistic setting, incorporating prior information about relationships between parameter values could substantially improve inference, as in Fig.~\ref{f:KDE} of the main text.

\begin{figure}
\center
\includegraphics[width=\linewidth]{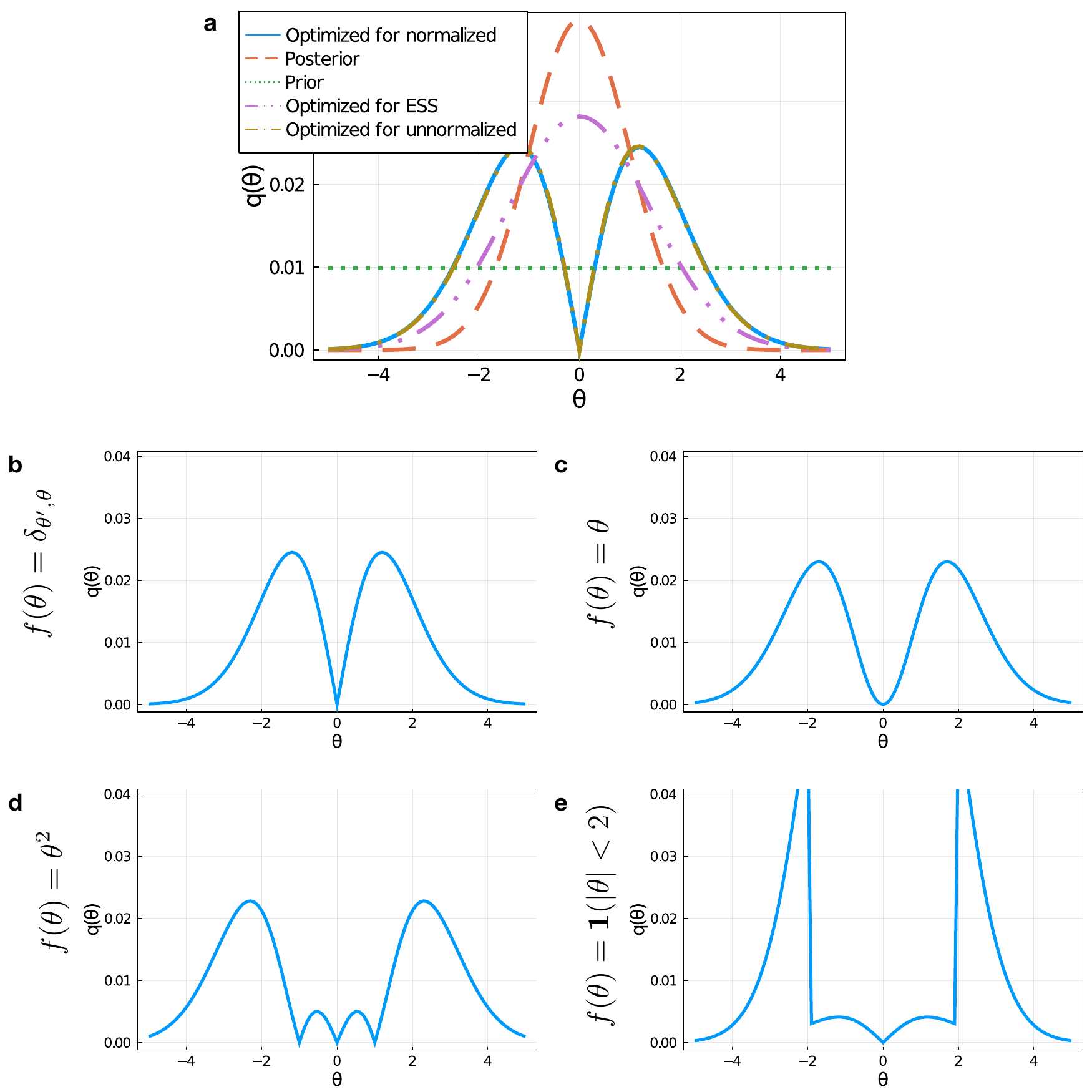}
\caption{
The optimal distribution of simulations depends strongly on the target function $f(\theta)$. Here $\Theta$ contains 101 evenly spaced points between $-5$ and $5$. 
The likelihood is $p(x^* |\theta) = e^{-\theta^2/2}$; the prior is uniform.
(a) Distributions of simulations optimized for criteria independent of $f(\theta)$. 
(b-e) Distributions optimized for the posterior MSE (b), mean (c), second moment (d), and $95\%$ confidence interval (e).
}
\label{f:discrete_gaussian_distributions}
\end{figure}

As before, we compare the effect of choosing simulation parameters from different distributions.
The prior, posterior, and optimal distributions for $\widehat{ESS}$ and unnormalized variance are all fixed by the problem setup (Fig.~\ref{f:discrete_gaussian_distributions}a).
The shape of the distribution optimized for estimating $\mathbb{E}_{p(\theta|x^*)}[f(\theta)]$, on the other hand depends strongly on $f$ (Fig.~\ref{f:discrete_gaussian_distributions}b-e).
Using the target $f(\theta) = \delta_{\theta', \theta}$, which is equivalent to targeting the normalized posterior, is nearly identical to targeting the unnormalized posterior here; the error in the normalizing constant is relatively small.

\begin{figure}
\center
\includegraphics[width=\linewidth]{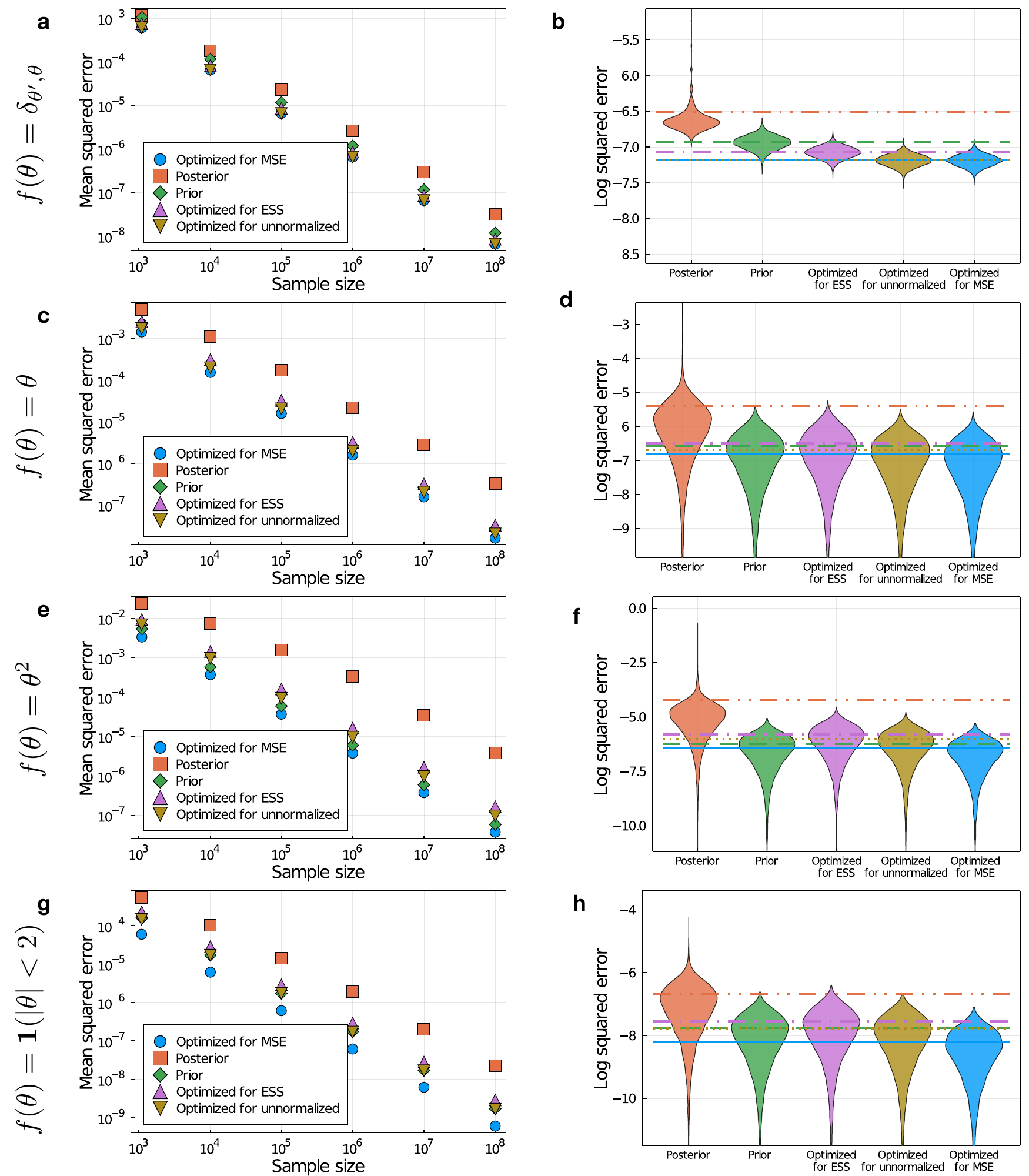}
\caption{
Optimizing the distribution of simulations for a specific target improves accuracy. As in Fig.~\ref{f:discrete_gaussian_distributions}, $\Theta$ contains 101 evenly spaced points between $-5$ and $5$. 
The likelihood is $p(x^* |\theta) = e^{-\theta^2/2}$; the prior is uniform.
The left column shows the MSE in the posterior (a), mean (c), second moment (e), and $95\%$ confidence interval (g) for each distribution of simulation parameters; the right column shows the distribution of squared errors across $10,000$ trials.
Here, unlike in Fig.~\ref{f:two_sample_score_comparison}, optimizing for the unnormalized posterior error gives approximately the same results as optimizing for the normalized posterior error via the function $f(\theta) = \delta_{\theta', \theta}$.
Each distribution was used to choose simulation parameters after the first 101, which where assigned uniformly to ensure every parameter value was used at least once.
We omitted inverse binomial sampling from comparison because it led to errors orders of magnitude higher than anything else.
}
\label{f:discrete_gaussian_errors}
\end{figure}

Each algorithm was initialized with one sample from each parameter value to ensure there was some information about every likelihood.
The resulting mean squared errors for four candidate functions (Fig.~\ref{f:discrete_gaussian_errors}, left column) all decrease at the same $1/N$ rate with varying scales.
Choosing parameters according to the posterior is relatively inefficient; optimizing either $\widehat{ESS}$ or the unnormalized posterior error performs better, though not optimally for all functions.
The relative efficiency of sampling from the prior depends significantly on the function to be estimated.
The largest efficiency gap occurs for the indicator function $f(\theta) = \mathbbm{1}(|\theta| < 2)$ corresponding approximately to a $95\%$ credible interval. Estimating that expectation accurately requires substantially more sampling in the higher-likelihood regions of the tails (Fig.~\ref{f:discrete_gaussian_distributions}e).

\subsection{Adaptive choice of parameters}
\label{app:discrete_adaptive}
The examples of Figs.~\ref{f:two_sample_MSE_simulations}-\ref{f:two_sample_score_comparison} and Fig.~\ref{f:discrete_gaussian_errors} show that, while optimizing for an approximate score like the effective sample size may not provide any improvement over rejection sampling, 
a more efficient targeted distribution of simulation parameters exists for any given function.
To reach that optimal efficiency, we need to approximate the optimal distribution of simulations without knowing the true posterior.
Fortunately, both unknowns in Eq.~\ref{e:discrete_optimal_n}, $p(x|\theta)$ and $\bar f$, can be adaptively estimated using likelihood estimates from past simulations.

Explicitly, we perform $M$ rounds of simulations starting with parameters chosen following the prior. 
In each round $m>1$, we compute the optimal $n_{i,m}$ using Eq.~\ref{e:discrete_optimal_n} based on the total budget of $mN/M$ simulations up to the current round, with likelihoods estimated using all previous simulations.
We then assign the $N/M$ simulations of round $m$ to bring the distribution of simulations performed as close as possible to the estimated optimum.
We do not attempt to optimize the number of rounds, instead fixing $M=16$ arbitrarily.
Though Eq.~\ref{e:discrete_optimal_n} is only valid in the limit $n_i\rightarrow\infty$, it can recommend $n_i =0$ for some parameters. 
To make sure $n_i$ is not too small for the asymptotic results to be relevant, we add $1/\sqrt{N}$ to the relative proportions in Eq.~\ref{e:discrete_optimal_n}, thereby approximately setting a floor of $\sqrt{N}$ simulations per parameter.

\begin{figure}
\includegraphics[width=\linewidth]{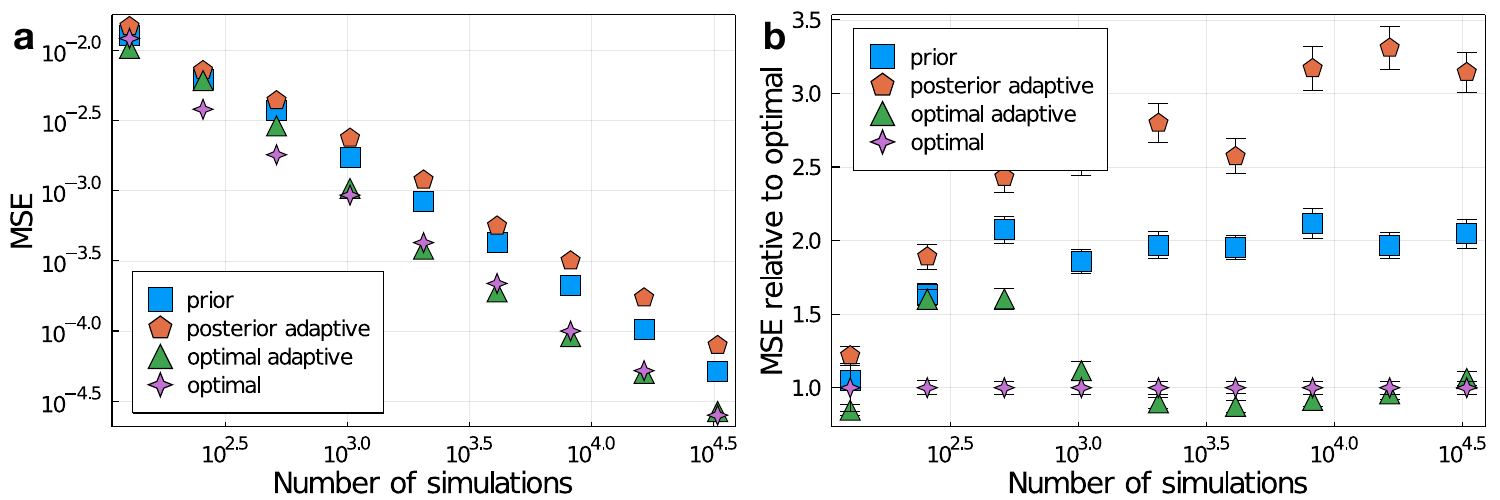}
\caption{
Adaptively targeting the optimal distribution can yield an approximately optimal MSE. 
Here, $\theta$ is an integer between 1 and 10, the likelihood is $e^{-(\theta - 5.5)^2/2}$, the prior is uniform, and the target function is $f(\theta) = \theta$.
(a) Mean of $\left(\mathbb{E}_{\hat p(\theta|x^*)}[\theta] - \mathbb{E}_{p(\theta|x^*)}[\theta]\right)^2$ across 1000 trials with four different sampling strategies:
sampling $\theta$ from the prior (blue squares), sampling $\theta$ from an adaptively-constructed approximation to the posterior (red pentagons), sampling $\theta$ from an adaptively-constructed approximation to Eq.~\ref{e:discrete_optimal_n} (green triangles), and sampling $\theta$ directly from Eq.~\ref{e:discrete_optimal_n} (purple stars).
The adaptive approaches used 16 rounds; for each round, the likelihood estimated from all previous simulations  was used to choose the next set of $N/16$ simulation parameters.
For all approaches, we assigned a minimum of $1/\sqrt{N}$ simulations to each parameter.
(b) MSE for the same experiment rescaled by the MSE with the optimal sampler.
Error bars  show empirical standard errors ignoring variance of the denominator; judging by the variation across $N$, these likely underestimate variability.
}
\label{f:discrete_adaptive}
\end{figure}

In Fig.~\ref{f:discrete_adaptive}, we compare three practically available options (sampling parameters from the prior, sampling parameters from an adaptively constructed approximation to the posterior, and choosing parameters from the above adaptive estimate of Eq.~\ref{e:discrete_optimal_n}) to sampling parameters from the optimal distribution.
For small numbers of simulations, there is not enough information to properly adapt the proposal, and hence no improvement over sampling from the prior.
For larger numbers of simulations, the adaptive approach matches the MSE of the optimal distribution.
In this example, adapting to the posterior performs poorly.

\begin{figure}
\includegraphics[width=1\linewidth]{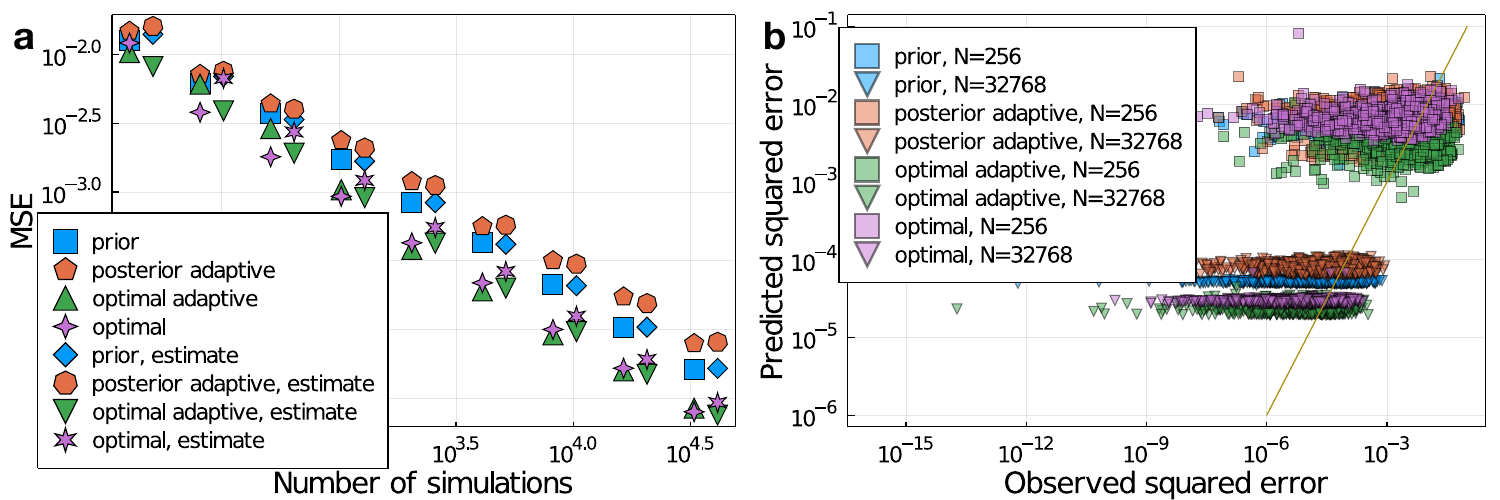}
\caption{
Eq.~\ref{e:discrete_asymptotic_variance} provides a rough estimate of the mean squared error in any posterior expectation estimate. 
As in Fig.~\ref{f:discrete_adaptive}, $\theta$ is an integer between 1 and 10, the likelihood is $e^{-(\theta - 5.5)^2/2}$, the prior is uniform, and the target function is $f(\theta) = \theta$.
(a) Mean of $\left(\mathbb{E}_{\hat p(\theta|x^*)}[\theta] - \mathbb{E}_{p(\theta|x^*)}[\theta]\right)^2$ across 1000 trials with four different strategies for choosing parameters:
sampling $\theta$ from the prior (blue squares), sampling $\theta$ from an adaptively-constructed approximation to the posterior (red pentagons), choosing $\theta$ from an adaptively-constructed approximation to Eq.~\ref{e:discrete_optimal_n} (green triangles pointing upward), and sampling $\theta$ directly from Eq.~\ref{e:discrete_optimal_n} (purple 4-pointed stars).
The remaining symbols are the mean across trials of variance estimates from Eq.~\ref{e:discrete_asymptotic_variance} using the same four strategies.
The adaptive approaches used 16 rounds; for each round, the likelihood estimated from all previous simulations  was used to choose the next set of $N/16$ simulation parameters.
For all approaches, we assigned a minimum of $1/\sqrt{N}$ simulations to each parameter.
The horizontal position of each MSE or estimate was moved slightly left or right respectively to minimize overlap.
(b)
Variance estimates from Eq.~\ref{e:discrete_asymptotic_variance} compared to $\left(\mathbb{E}_{\hat p(\theta|x^*)}[\theta] - \mathbb{E}_{p(\theta|x^*)}[\theta]\right)^2$ for individual trials from the left panel.
The brown line indicates correct prediction.
}
\label{f:discrete_adaptive_error_estimation}
\end{figure}

A final desirable piece of information is the level of uncertainty in our estimate of $\bar f$.
Like the distribution of optimal samples, the variance approximation from Eq.~\ref{e:discrete_asymptotic_variance} can be estimated from the algorithm output by substituting $\hat p_i^*$ for $p_i^*$. 
That estimation can be done for any function, regardless of how the simulation parameters were chosen.
For the example in Figs.~\ref{f:discrete_adaptive} and~\ref{f:discrete_adaptive_error_estimation}, this adaptive error estimate approximately matches the mean squared error of different sampling strategies, with higher accuracy for larger $N$ (Fig.~\ref{f:discrete_adaptive_error_estimation}a).
This masks considerable variation in $\left(\mathbb{E}_{\hat p(\theta|x^*)}[\theta] - \mathbb{E}_{p(\theta|x^*)}[\theta]\right)^2$ across trials (Fig.~\ref{f:discrete_adaptive_error_estimation}b).
Importantly, high error in the posterior estimate would make the variance estimate similarly unreliable; the uncertainty calculation can only be trusted if the posterior approximation is good, as it is in Fig.~\ref{f:discrete_adaptive_error_estimation}.

\section{Variance if sampling $\theta$}
\label{app:sampling_parameters}
\subsection{Independent sampling}
\label{app:independent_sampling}
In deriving Eq.~\ref{e:discrete_asymptotic_variance}, we assumed that the simulation parameters were chosen deterministically or, equivalently, via stratified sampling with $|\Theta|$ strata.
 Often, however, parameters are instead sampled independently from an importance distribution $q(\theta)$. 
Independent sampling leads to a different asymptotic variance, which we compute here without assuming the parameters are discrete.
The posterior expectation estimate in Eq.~\ref{e:rejection_sampling_expectation} can be rewritten as
\begin{equation}
\mathbb{E}_{\hat p(\theta|x^*)} \left[ f(\theta) \right]
= 
\frac{ \frac{1}{N} \sum_{i = 1}^N f(\theta_i) \frac{p(\theta_i)}{q(\theta_i)} \mathbbm{1}(x_i = x^*)}{\frac{1}{N} \sum_{i = 1}^N  \frac{p(\theta_i)}{q(\theta_i)} \mathbbm{1}(x_i = x^*)}
\label{e:importance_sampling_expectation}
\end{equation}
In the numerator,
\begin{align}
\mathbb{E}_{\theta \sim q, x \sim p(x|\theta)} \left[
f(\theta) \frac{p(\theta)}{q(\theta)} \mathbbm{1}(x= x^*)
\right]
&= p(x^*) \bar f.
\end{align}
In the denominator,
\begin{align}
\mathbb{E}_{\theta \sim q, x \sim p(x|\theta)} \left[
\frac{p(\theta)}{q(\theta)} \mathbbm{1}(x= x^*)
\right]
&= p(x^*).
\end{align}
We then use the delta method, specifically Eq.~\ref{e:delta_method_reorganized} with $\mu_R/\mu_S = \bar f$, to find
\begin{align}
\var\left(\mathbb{E}_{\hat p(\theta|x^*)} \left[ f(\theta) \right]
\right)
&\approx \frac{1}{p(x^*)^2} \var
\left(
\frac{1}{N} \sum_{i=1}^N \left(f(\theta_i) - \bar f\right)
\frac{p(\theta_i)}{q(\theta_i)} \mathbbm{1}[x_i = x^*]
\right)
\\
& = \frac{1}{Np(x^*)^2} \var
\left( \left(f(\theta_i) - \bar f\right)
\frac{p(\theta_i)}{q(\theta_i)} \mathbbm{1}[x_i = x^*]
\right)
\\
& = \frac{1}{Np(x^*)^2} 
\mathbb{E}_{\theta \sim q, x \sim p(x|\theta)}\left[
\left( \left(f(\theta_i) - \bar f\right)
\frac{p(\theta_i)}{q(\theta_i)} \mathbbm{1}[x_i = x^*]
\right)^2
\right].
\end{align}
Substituting in $\mathbb{E}_{x\sim p(x|\theta)}\left[\mathbbm{1}[x_i = x^*] \right] = p(x^*|\theta)$ and writing out the integral for the expectation over $\theta$, we are left with
\begin{equation}
\var\left(\mathbb{E}_{\hat p(\theta|x^*)} \left[ f(\theta) \right]
\right)
\approx \frac{1}{Np(x^*)^2} \int (f(\theta) - \bar f)^2 \frac{p(\theta)^2}{q(\theta)} p(x^* |\theta) d\theta.
\label{e:independent_sampling_variance}
\end{equation}

Sampling from either the posterior or the prior leads to a simpler variance expression.
If $q(\theta) = p(\theta|x^*) = p(x^*|\theta)p(\theta)/p(x^*)$,
\begin{align}
\var\left(\mathbb{E}_{\hat p(\theta|x^*)} \left[ f(\theta) \right]
\right)
&\approx \frac{1}{Np(x^*)^2} \int (f(\theta) - \bar f)^2 \frac{p(x^*)p(\theta)^2}{p(x^*|\theta) p(\theta)} p(x^* |\theta) d\theta
\\
& = \frac{1}{Np(x^*)} \int (f(\theta) - \bar f)^2 p(\theta)  d\theta,
\label{e:posterior_sampling_variance}
\end{align}
while if $q(\theta) = p(\theta)$
\begin{align}
\var\left(\mathbb{E}_{\hat p(\theta|x^*)} \left[ f(\theta) \right]
\right)
&\approx \frac{1}{Np(x^*)^2} \int (f(\theta) - \bar f)^2 \frac{p(\theta)^2}{p(\theta)} p(x^* |\theta) d\theta
\\
& = \frac{1}{Np(x^*)} \int (f(\theta) - \bar f)^2 p(\theta|x^*)  d\theta.
\label{e:prior_sampling_variance}
\end{align}
That is, in addition to a shared factor $p(x^*)^{-1}$, the posterior sampling variance depends on the prior expectation of $\left(f(\theta) - \bar f\right)^2$ while the prior sampling variance depends on the posterior expectation of $\left(f(\theta) - \bar f\right)^2$. 

A simple $f$ to consider is an indicator function for a $50\%$ credible interval, where $f(\theta) - \bar f = \pm 0.5$ everywhere. Substituting into either Eq.~\ref{e:posterior_sampling_variance} or Eq.~\ref{e:prior_sampling_variance} gives the same approximate variance $\left(4Np(x^*)\right)^{-1}$.
Often, however, the prior will have more mass in regions where $(f(\theta) - \bar f)^2$ is large; in that case, the variance with posterior sampling is larger than the variance with prior sampling. 

With small sample sizes, the error with posterior sampling may often be smaller than expected.
A substantial part of the posterior MSE comes from low-probability events, where some $\theta_i$ with low $p(x^*|\theta)$ nevertheless yields $x_i = x^*$ and is included with high weight $p(\theta|x^*)^{-1}$.
Our empirical estimates of the posterior $\widehat{ESS}$ and MSE in Figs.~\ref{f:continuous_simple} and~\ref{f:discrete+continuous}  suggest worse relative performance with higher $N$, likely due to more regularly seeing these high-weight outliers.
Approaches to address this issue, such as resampling~\cite{Sisson2020}, are beyond the scope of this paper.

Neither prior sampling nor posterior sampling is optimal.
By a similar argument to Section~\ref{app:optimal_MSE_distribution}, 
optimizing Eq.~\ref{e:independent_sampling_variance} over $q(\theta)$ yields
\begin{equation}
q(\theta) \propto p(\theta) \sqrt{p(x^*|\theta)}|f(\theta) - \bar f|.
\label{e:independent_sampling_optimal_n_appendix}
\end{equation}
Compared to the optimal distribution for deterministically chosen discrete parameters, this lacks the factor $\sqrt{1 - p(x^*|\theta)}$.
The difference, in the discrete case, comes from weighting by the importance distribution $q(\theta_i)$ in Eq.~\ref{e:importance_sampling_expectation}, i.e. weighting by the expected rather than the actual number of times $\theta_i$ was used for simulations.
If $p(x^*|\theta)$ is close to 1, $n_i^*/n_i$ can be an accurate estimate of  $p(x^*|\theta)$ with small $n_i$; $n_i^*/\mathbb{E}[n_i]$ is only accurate if $n_i/\mathbb{E}[n_i]$ concentrates near 1, which happens if $\mathbb{E}[n_i]$ is large.

\subsection{Stratified sampling}
\label{app:stratified}

An alternative to independently sampling $\theta_i$ from an importance distribution $q(\theta)$ is \textit{stratified sampling}, where we partition the parameter space $\Theta$ into $K$ disjoint strata $\{\Omega_k\}$,  and for each $k$ sample $n_k$ simulation parameters $\{\theta_{k,i}\}$ from an importance distribution $q_k(\theta)$ with support $\Omega_k$.
To account for the stratification, we adjust the weights in Eq.~\ref{e:importance_sampling_expectation} to
\begin{equation}
\mathbb{E}_{\hat p(\theta|x^*)} \left[ f(\theta) \right]
= 
\frac{  \sum_{k = 1}^K \frac{1}{n_k}\sum_{i = 1}^{n_k} f(\theta_{k,i}) \frac{p(\theta_{k,i})}{q_k(\theta_{k,i})} \mathbbm{1}(x_{k,i} = x^*)}{  \sum_{k = 1}^K \frac{1}{n_k}\sum_{i = 1}^{n_k}  \frac{p(\theta_{k,i})}{q_k(\theta_{k,i})} \mathbbm{1}(x_{k,i} = x^*)}
\equiv
\frac{\sum_{k=1}^K R_k}{\sum_{k=1}^K S_k} \equiv \frac{R}{S}.
\end{equation}
Here $R_k$ is an unbiased estimate of the posterior integral over the stratum $\Omega_k$:
\begin{align}
\mathbb{E}[R_k] &= \int_{\Omega_k} f(\theta) \frac{p(\theta)}{q_k(\theta)} p(x^*|\theta) q_k(\theta) \dd\theta
\\
&= p(x^*) \int_{\Omega_k} f(\theta)  p(\theta|x^*) \dd \theta.
\end{align}
Because the $\Omega_k$ form a partition of $\Theta$,
\begin{equation}
\mathbb{E}\left[\sum_k R_k\right] = \sum_k p(x^*) \int_{\Omega_k} f(\theta)  p(\theta|x^*) \dd \theta = p(x^*) \bar f.
\end{equation}
Similarly, $\mathbb{E}[S] = p(x^*)$.

Using the delta method in the form of Eq.~\ref{e:delta_method_reorganized},
\begin{align}
\var\left(
\mathbb{E}_{\hat p(\theta|x^*)} \left[ f(\theta) \right]
\right)
&\approx
\frac{1}{p(x^*)^2}\var\left(\sum_{k=1}^K \left(R_k - \bar f S_k\right)
\right)
\\
& = \frac{1}{p(x^*)^2}\sum_{k=1}^Kn_k ^{-1}\var\left( (f(\theta_{k,i}) - \bar f) \frac{p(\theta_{k,i})}{q_k(\theta_{k,i})} \mathbbm{1}(x_{k,i} = x^*)
\right).
\label{e:stratified_sampling_variance}
\end{align}
Equation~\ref{e:stratified_sampling_variance} generalizes our result for the asymptotic variance with discrete parameters (Eq.~\ref{e:discrete_asymptotic_variance}). 
The discrete case can be considered as stratified sampling with each stratum containing a single parameter value, which implies $f(\theta_{k,i})$ and $p(\theta_{k,i})$ are both constant and $q_k(\theta_{k,i}) = 1$.

Qualitatively, we expect similar behavior for fine stratification, where 
$f(\theta)$, $p(\theta)$, and $p(x^*|\theta)$ are approximately constant on $\Omega_k$ because $\Omega_k$ is small and 
$q_k$ may be chosen to be close to uniform over $\Omega_k$.
Then we may simplify the approximate variance:
\begin{align}
\var\left(
\mathbb{E}_{\hat p(\theta|x^*)} \left[ f(\theta) \right]
\right)
&\approx
 \frac{1}{p(x^*)^2}\sum_{k=1}^Kn_k ^{-1}(f(\theta_k) - \bar f)^2 \frac{p(\theta_{k})^2}{q_k(\theta_{k})^2}\var\left(  \mathbbm{1}(x_{k} = x^*)
\right)
\\
& \approx
\frac{1}{p(x^*)^2}\sum_{k=1}^Kn_k ^{-1}V(\Omega_k)^2(f(\theta_k) - \bar f)^2 p(\theta_k)^2p(x^*|\theta_k) (1 - p(x^*|\theta_k)),
\label{e:stratified_sampling_simplified_variance}
\end{align}
where $\theta_k$ is an arbitrary point in $\Omega_k$, $V(\Omega_k)$ is the volume of stratum $k$, and we used $q_k(\theta_k)\approx V(\Omega_k)^{-1}$.
Having a constant number of samples per stratum would be optimal if
\begin{equation}
V(\Omega_k) \propto \left( p(\theta_k) |f(\theta_k) - \bar f| \sqrt{p(x^*|\theta_k) (1 - p(x^*|\theta_k)}\right)^{-1}.
\end{equation}
This can be achieved by dividing $\Theta$ into strata with equal probability according to the distribution 
\begin{equation}
q(\theta) \propto p(\theta_k) |f(\theta_k) - \bar f| \sqrt{p(x^*|\theta_k) (1 - p(x^*|\theta_k)},
\label{e:stratified_optimal_n}
\end{equation}
which was optimal for the discrete case. In the example in Fig.~\ref{f:KDE}, applying Eq.~\ref{e:stratified_optimal_n} outperformed any of the other sampling strategies we tried.

The above argument is informal: we have not carefully considered the relationship between the stratification and variation in $f(\theta)$, $p(\theta)$, and $p(x^*|\theta)$.
A rigorous derivation of a usable approximation to Eq.~\ref{e:stratified_sampling_variance}, including bounds on the approximation error, would be worthwhile but is beyond the scope of this paper.

\section{Likelihood estimation with kernel regression}
\label{app:kernel_regression}

Here we give further details of the kernel regression approach we used in Fig.~\ref{f:KDE}. 
We start from a set of $N$ simulation parameters  and outputs $(\theta_i, x_i)$ and assign weights $w_i = \mathbbm{1}[x_i = x^*]$.
Then for any $\theta$, we use a Nadaraya-Watson estimator of the likelihood:
\begin{equation}
\hat p(x^*|\theta) = \frac{\sum_{i=1}^N w_iK_h(\theta - \theta_i)}{\sum_{i=1}^N K_h(\theta - \theta_i)},
\end{equation}
where $K_h(\alpha) = h^{-1} K\left(h^{-1}\alpha\right)$ is a kernel with bandwith $h$.
We chose this formula, rather than the kernel estimate used in \cite{Blum2010} that does not normalize based on $K_h(\theta - \theta_i)$, so that our estimator does not rely on $\theta_i$ being sampled from the prior.
Function expectations can be computed by integrating
\begin{equation}
\mathbb{E}_{\hat p (\theta|x^*)} \left[f(\theta)\right]
= \frac{\int f(\theta) \hat p(x^*|\theta) p(\theta) \dd\theta}{\int \hat p(x^*|\theta) p(\theta) \dd\theta},
\end{equation}
which we did with Gauss-Kronrod quadrature.
We chose a Gaussian kernel $K(\alpha) = \exp(-\alpha^2/2)$, set $h = N^{-1/2} $, and used 20 quadrature points. These choices were sufficient to consistently improve on the empirical average in Eq.~\ref{e:rejection_sampling_expectation} but were not optimized.

\section{ABC-SMC with particles from all rounds}
\label{app:ABC-SMC_keeping_all_rounds}

ABC-SMC methods~\cite{DelMoral2012, Filippi2013,Sisson2007, Toni2009} 
use $K$ rounds of importance sampling with adaptively constructed proposal distributions 
to produce a weighted set of samples from an approximate posterior.
The process begins with a round of rejection sampling from the prior $p(\theta) = q_1(\theta)$.
In each round $k>1$, each new simulation parameter $\theta_{k,i}$ is selected by first sampling $\tilde \theta_{k,i}$ from the weighted samples $\{(\theta_{k-1, j}, w_{k-1, j})\}$ from the previous round and then perturbing $\tilde \theta_{k,i}$.
Implicitly, then, each round draws $N_k$ parameters $\theta_{k,i}$ from an importance distribution
\begin{equation}
q_k(\theta) = W_{k-1}^{-1}\sum_j w_{k-1,j} q(\theta_{k,i}|\theta_{k-1, j})
\end{equation}
where $W_{k-1}^{-1} = \sum_j w_{k-1,j}$ is the sum of the weights from round $k-1$ and $q(\theta_{k,i}|\theta_{k-1, j})$ is a perturbation kernel to be specified.
After a stopping criterion is reached, which could be a fixed number of simulations performed or a fixed number of simulations accepted, particles where the simulated data is farther than $\epsilon_k$ from $x^*$ are discarded and the remaining $N_{acc, k}$ used with weights $p(\theta)/q_k(\theta)$ as the base for the next importance distribution $q_{k+1}(\theta)$. The threshold $\epsilon_k$ may be fixed at the start or chosen so that a desired proportion of particles are accepted.

In the standard version of ABC-SMC, the output is the weighted set of accepted particles from the final round.
Simulations from prior rounds are ignored.
An alternative is to include any particle $\theta_{k,i}$ where the discrepancy $\Delta(x_{k,i}, x^*) < \epsilon_K$, even if $k$ was not the final round $K$.
This effectively changes the importance distribution for the output from $q_K$ to 
\begin{equation}
q(\theta) = \frac{1}{N}\sum_k N_k q_k (\theta)
\label{e:ABC_SMC_keep_proposal}
\end{equation}
and the number of simulations that could be output from $N_K$ to $N = \sum_k N_k$.
The standard importance sampling weight for $\theta_{k,i}$ would then be  $\frac{p(\theta_{k,i})}{q(\theta_{k,i})}$.
Alternatively, we can consider each round as giving an independent estimate of $\bar f$ with the final threshold $\epsilon_K$, 
\begin{equation}
\bar f_k \equiv \mathbb{E}_{\hat p_k(\theta|x^*)} \left[f(\theta)\right] = \frac{\sum_i \frac{p(\theta_{k,i})}{q_k(\theta_{k,i})}\mathbbm{1}\left[\Delta (x_{k,i}, x^*) < \epsilon_K \right] f(\theta_{k,i})}
{\sum_i \frac{p(\theta_{k,i})}{q_k(\theta_{k,i})}\mathbbm{1}\left[\Delta (x_{k,i}, x^*) < \epsilon_K \right]},
\end{equation}
and combine the estimates with round weights $\alpha_k$,
\begin{equation}
\mathbb{E}_{\hat p(\theta|x^*)} \left[f(\theta)\right] = \frac{\sum_k \alpha_k \bar f_k}{\sum_k \alpha_k},
\label{e:ABC_SMC_heuristic_keep_weights}
\end{equation}
thereby giving each accepted particle weight $\alpha_k p(\theta_{k,i})/q_k(\theta_{k,i})$.
Ideally, $\alpha_k$ would be inversely proportional to the variance of $\bar f_k$ given by Eq.~\ref{e:independent_sampling_variance}.
In our examples, we found setting $\alpha_k$ equal to the effective sample size for round $k$ with threshold $\epsilon_K$ often sufficient to substantially improve on the standard ABC-SMC choice $\alpha_k = 0$ for $k<K$.

We use the latter weights $\alpha_k p(\theta_{k,i})/q_k(\theta_{k,i})$ rather than $p(\theta_{k,i})/q(\theta_{k,i})$ to avoid slow computations of $q(\theta)$ in Eq.~\ref{e:ABC_SMC_keep_proposal}, though we have no reason to believe either choice of weights is optimal.
Related strategies for efficiently reusing earlier simulations have been proposed with other algorithms~\cite{Lueckmann2017, Papamakarios2019}.

In the remainder of this section, we compare the effect of accepting particles from all rounds with the effect of two other algorithm choices, the perturbation kernel and the choice of thresholds $\epsilon_k$.
We use an example from~\cite{Filippi2013}, 
where Filippi et al. investigated the effect of the choice of perturbation kernel on the acceptance rate at each round.

\begin{figure}
\includegraphics[width=\linewidth]{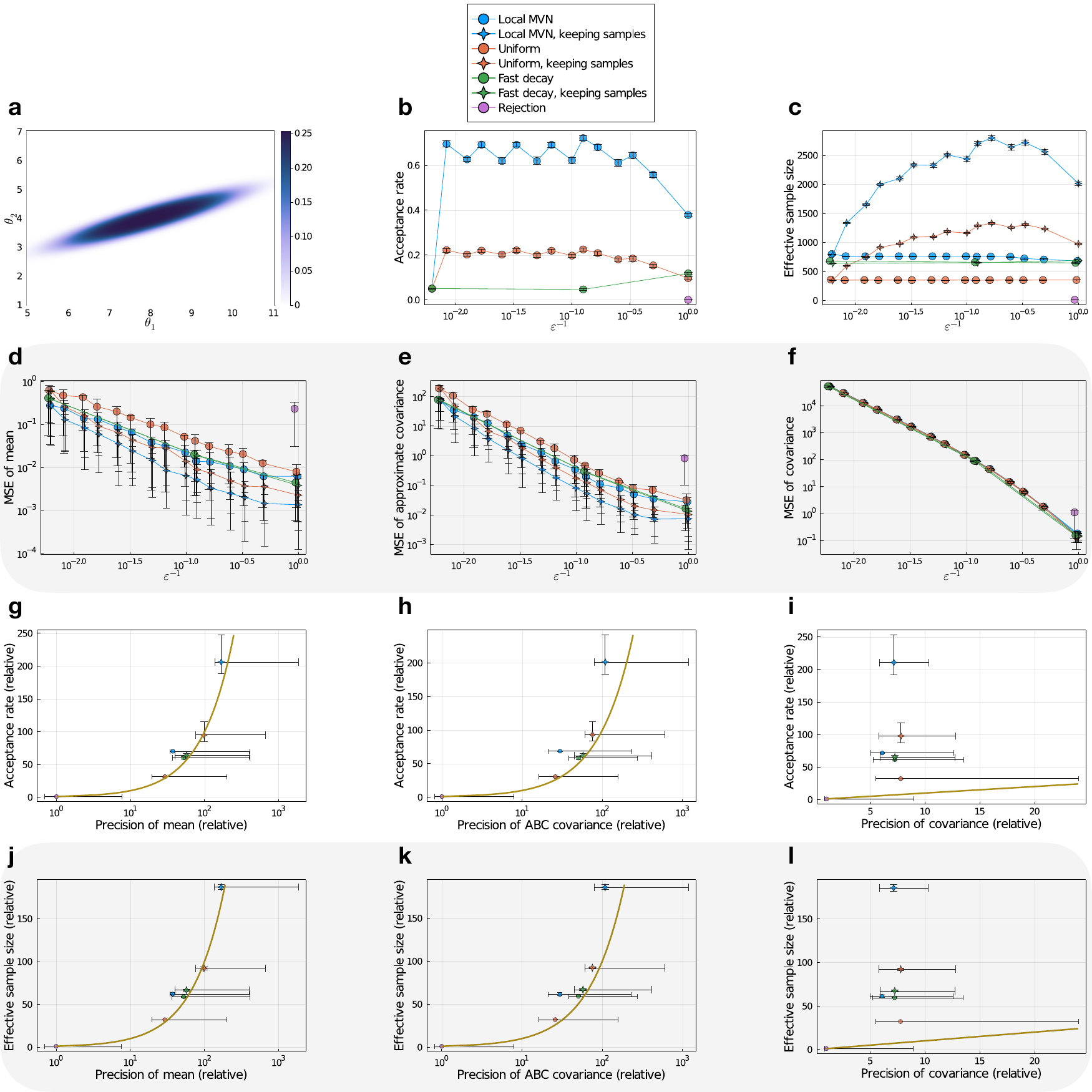}
\caption{The acceptance rate and $\widehat{ESS}$ imperfectly measure accuracy in an example considered by Filippi et al.~\cite{Filippi2013}.
The true posterior (a) is an ellipsoid centered on $(8, 4)$.
In (b), we replicate the evaluation done in~\cite{Filippi2013}: decreasing $\epsilon$ slowly with a local multivariate normal kernel (MVN, blue) yields higher per-round acceptance rates $N_{acc,k}/N_k$ than a uniform kernel (red), multivariate normal kernel with rapidly decaying $\epsilon$ (green), or direct rejection sampling (purple).
(c) The MVN kernel likewise gives the highest $\widehat{ESS}$ of the standard ABC-SMC approaches, though for the slow schedules $\widehat{ESS}$ can be substantially increased by including particles from all rounds (stars).
(d-f) The MSE (averaged over 100 trials) in estimating the mean (d), variance of the ABC approximate posterior (e), and variance of the true posterior (f) improves with decreasing $\epsilon$ at similar rates for all methods.
All ABC-SMC algorithms perform equivalently in estimating the true posterior variance (f), where bias from finite $\epsilon$ is significant.
(g-l)
To compare scores, we plot the mean final acceptance rate $N_{acc,K}/N$ (g-i) and $\widehat{ESS}$ (j-l) for each approach against the precision in estimating the same three targets as (d-f), with all scores divided by the corresponding score for rejection sampling.
The brown line shows equal improvement in both plotted scores, ignoring the uncertainty in the baseline rejection sampling score.
All approaches used around $34,000$ simulations in total. All error bars show 25th and 75th percentiles.
}
\label{f:ABC_SMC}
\end{figure}

We consider two real parameters $\theta_1$ and $\theta_2$, with independent uniform priors on $[-50, 50]$.
The likelihood for the continuous data $y$ is $p(y|\theta)=\mathcal{N}((\theta_1 - 2\theta-2)^2 + (\theta_2 - 4)^2, 1)$; $y^*=0$ is observed.
Using a decreasing sequence of thresholds $\epsilon_i$ effectively creates discrete observations $x^*_{\epsilon_i} = \mathbbm{1}(|y| <\epsilon_i)$.

In addition to the acceptance rates originally presented, we compute effective sample sizes and the mean squared error of posterior means and variances.
The ground truth can be computed analytically, which we do in Section~\ref{app:filippi_posterior}.
For each approach that we compare, we have a final tolerance $\epsilon = 1$ and tune the target number of acceptances per round so that $N \approx 34,000$ simulations are used in total.

Figure~\ref{f:ABC_SMC}b replicates the evaluation of acceptance rates from~\cite{Filippi2013}, finding as Filippi et al. did that a local multivariate normal  perturbation kernel with slow decrease of $\epsilon_i$ gives the highest per-round acceptance rates. 
Rapidly shrinking the tolerance significantly impacts the acceptance rate, while rejection sampling only accepts 11 samples out of $\sim 34000$.
When comparing effective sample sizes (Fig.~\ref{f:ABC_SMC}c), the change in $\epsilon$ schedule makes a much smaller difference.

In Fig.~\ref{f:ABC_SMC}d-l, we look at the accuracy of estimation of the mean (left column), ABC posterior covariance ignoring bias due to finite $\epsilon_i$ (ABC covariance, center column), and true posterior covariance (right column).
Rejection sampling performs the worst and accepting particles from all rounds improves estimates, but the changed $\epsilon$ schedule does not matter much.

Because of the symmetry of the true posterior, all mean estimates are unbiased.
The bias in the covariance, however, is significant enough that a rejection sampling estimate for $\epsilon = 1$, which had 11 accepted samples, outperforms the ABC-SMC estimate for $\epsilon = 2$ with $\widehat{ESS} \approx 2500$ (Fig.~\ref{f:ABC_SMC}f).
This is in contrast to the mean and ABC covariance, where the last few rounds of reducing $\epsilon$ make little difference.
For more discussion of the scale of bias from finite $\epsilon$, we refer readers to~\cite{Barber2015}.

The overall acceptance rate $N_{acc,K}/N$ (Fig.~\ref{f:ABC_SMC}g-i) and the effective sample size (Fig.~\ref{f:ABC_SMC}j-l) are correlated with the improvement of ABC-SMC relative to rejection sampling, but the mismatch between scores is on the same scale as the differences between algorithms.
Compared to rejection sampling, averaged over our trials ABC-SMC with a local MVN kernel and the $\epsilon$ schedule used by Filippi et al. had 60 times higher effective sample size, but only 40 or 15 times higher precision in estimating the mean or ABC covariance respectively.

\subsection{Deriving the true posterior}
\label{app:filippi_posterior}

In this section we analytically compute the true posterior for the example in Fig.~\ref{f:ABC_SMC}.
The only approximation we make is to use an improper uniform prior on $[-\infty, \infty]^2$ rather than the broad but finite uniform prior on $[-50,50]^2$ considered in~\cite{Filippi2013}.
Up to a normalizing constant $Z$, we have
\begin{equation}
p(\theta|y = 0)= \frac{1}{Z}\exp\left[- \frac{1}{2}\left((\theta_1 - 2\theta_2)^2 + (\theta_2 -4)^2 \right)^2\right].
\end{equation}

To simplify later calculations, we define new variables  $u_1 = \theta_1 - 2\theta_2$ and $u_2 = \theta_2 - 4$; equivalently,
\begin{equation}
\ve{u} \equiv\begin{pmatrix}
u_1 \\u_2
\end{pmatrix}
=
\begin{pmatrix}
1 & -2 \\ 0 & 1
\end{pmatrix}
\begin{pmatrix}
\theta_1 -8 \\ \theta_2 - 4
\end{pmatrix} \equiv J \left(\theta - \mu_\theta\right)
\end{equation}
with $\mu_\theta = (8, 4)^\top$.
Applying a change of variables from $\ve{\theta}$ to $\ve{u}$,
\begin{equation}
p(\ve{u}| x) = \frac{1}{|J|Z} \exp\left[-\frac{1}{2}\left(u_1^2 + u_2^2\right)^2\right].
\end{equation}
The determinant of the Jacobian can be dropped, as $|J| = 1$.
We can now integrate in polar coordinates to find the normalizing constant:
\begin{align}
Z &= \int\int \exp\left[-\frac{1}{2}\|\ve{u}\|^4\right] \dd\ve{u}
\\
& = \int\int\exp\left[-\frac{1}{2}r^4\right] r\dd r \dd\phi
\\
& = 2\pi \int_0^\infty r\exp\left[-\frac{r^4}{2}\right] \dd r
\\
&= 2\pi \int_0^\infty \exp\left[-\frac{(r^2)^2}{2}\right] \frac{1}{2}\dd (r^2)
\\
& = \frac{\pi}{2} \int_{-\infty}^\infty \exp\left[-\frac{a^2}{2}\right] \dd a
\\
& = \frac{\pi}{2} \sqrt{2\pi} = \sqrt{\pi^3/2}.
\end{align}

In order to evaluate ABC-SMC algorithms, we need the mean and variance of the true posterior.
Because the distribution is symmetric in $\ve{u}$, the mean of $\ve{u}$ is $(0,0)$ and the mean of $\theta$ is $\mu_\theta = (8,4)$.
Moreover, the distribution is symmetric in $u_1$ for any $u_2$, so $\mathbb{E}[u_1 u_2] = 0$; the covariance is diagonal.
Again by symmetry, $\mathbb{E}[u_1^2] = \mathbb{E}[u_2^2]$.
The symmetry arguments leave one term to be calculated with an integral:
\begin{align}
\mathbb{E}[u_1^2] & = \frac{1}{Z} \int\int u_1^2 \exp\left[-\frac{1}{2}\|\ve{u}\|^4\right] \dd\ve{u}
\\
& = \frac{1}{Z}  \int\int r^2\cos(\phi)^2\exp\left[-\frac{1}{2}r^4\right] r\dd r \dd\phi
\\
& = \frac{1}{Z}  \left(\int_0^{2\pi} \frac{1}{2}\left( 1 - \cos(2\phi)\right) \dd\phi\right)\left(\int_0^\infty r^3 \exp\left[-\frac{r^4}{2}\right] dr\right)
\\
& = \frac{1}{Z} \pi \left( -\frac{1}{2}\exp\left[-\frac{r^4}{2}\right]\right)\bigg|^\infty_0
\\
& = \frac{\pi}{2Z}.
\end{align}

Substituting in $Z = \sqrt{\pi^3/2}$, $\mathbb{E}[u_1^2] = (2\pi)^{-1/2}$. Then using $\theta = J^{-1} \ve{u} + \mu_\theta$.
\begin{align}
\cov(\theta) &= \mathbb{E}[(\theta-\mu_\theta)(\theta-\mu_\theta)^\top] = J^{-1}\mathbb{E}[\ve{u}\ve{u}^\top]J^{-\top} = (2\pi)^{-1/2} J^{-1}J^{-\top}
\\
& = (2\pi)^{-1/2}
\begin{pmatrix}
5 & 2
\\
2 & 1
\end{pmatrix}.
\end{align}
\end{document}